\documentclass[12pt,preprint,aps,showkeys]{revtex4}
\usepackage{amsmath}
\usepackage{amsfonts}
\usepackage{amssymb}
\usepackage{natbib}
\usepackage{graphicx}
\usepackage{epstopdf}
\usepackage{subfigure}
\usepackage{mathrsfs}
\usepackage{dcolumn}
\usepackage{bm}
\usepackage[section]{placeins}
\usepackage{hyperref}
\usepackage{xcolor}
\begin{document}
\title{Mutual information for particle pair  and its application to diagnose chaos in curved spacetime}
\author{Wenfu Cao$^{1}$}
\email{202411100001@stu.ujn.edu.cn}
\author{Yang Huang$^{2}$}
\email{sps\_Huangy@ujn.edu.cn}
\author{Hongsheng Zhang$^{3}$}
\email{sps\_zhanghs$@$ujn.edu.cn}

 \affiliation{School of Physics and Technology, University of Jinan,
336 West Road of Nan Xinzhuang, Jinan, Shandong 250022, China}

\begin{abstract}
  We propose the concept of mutual information for particle pair (MIPP) in curved spacetime, and show that MIPP has potential to be a proper chaos indicator. We tested this method in
  the Schwarzschild and Kerr spacetime and compared it with the fast Lyapunov indicator. The results show that the MIPP
 effectively identify orbital states and demonstrates prominent performance in recognizing transitions between orbital states. Our result show that information theory
 significantly deepen our understanding of dynamics of few-body system.
\end{abstract}
\maketitle
\section{Introduction}
The concept of chaos originated in the Poincar\'{e} era but still lacks a unified and generally accepted definition.
Here, we adopt a definition of chaos by Devaney \cite{Devaney:1990}: Let $V$ be a set, and $F: V\rightarrow V$ be a dynamical system.
If the following conditions are satisfied: (1) $F$ exhibits exponential sensitivity to initial conditions;
(2) $F$ is topologically transitive; (3) Periodic points are dense in $V$, then $F$ is called chaotic.
Although definitions of chaos vary literature by literature, they generally highlight several common characteristics of chaotic systems: exponential sensitivity to initial
conditions, deterministic randomness, unpredictability, and aperiodicity. Recent research has also shown that chaotic systems often exhibit significant entropy fluctuations
\cite{Cao:2024bjk}.

 Physical system that is non-integrable or whose integrability is disrupted by background perturbations typically exhibit chaotic behavior, and such system is widely
regarded as a major topic in classical nonlinear dynamics. With the development of high-precision numerical methods and the use of advanced computers, people gradually sharpen
the understanding of chaotic systems.

The theoretical foundation of chaos began with the introduction of the Kolmogorov-Arnold-Moser (KAM) theory of invariant torus in the 1950s-1970s.
Subsequently, a series of numerical explorations, such as Lorenz's weather system simulation \cite{Lorenz:1963yb}, H$\acute{e}$non and Heiles' study on the existence of the
third integral \cite{Henon1964}, and Sussman and Wisdom's discovery of chaotic motion in Pluto's orbit \cite{Gerald:1988}, demonstrated the achievements of chaos dynamics in
celestial mechanics.

The application of chaos theory has now been extended to relativistic systems, with substantial research on chaotic geodesic motion in modified gravity theory
\cite{Chen:2016tmr,
Li:2018wtz, Zhang:2023lrt, Kopacek:2010yr, Cao:2024ihv, Hu:2021gwd, Yi:2020shw, Sun:2021oxg, Sun:2021ndd}. One of the latest studies in this field explores the chaotic
dynamics of stellar-mass compact objects being accreted by supermassive black holes. The resulting motion is
typically non-integrable, and this non-integrability leads to chaotic behavior of the accreting objects. The chaotic nature of this motion leaves distinctive imprints in the
gravitational wave signals \cite{Destounis:2021mqv}. Meanwhile, chaos identification indicators have been continuously refined and developed,
including methods such as the Poincar\'{e} map, Lyapunov exponent \cite{Tancredi:2001,Wu:2003pe}, local Lyapunov exponent and its spectral distribution
\cite{Voglis1994}, fast
Lyapunov indicators \cite{Wu:2006rx}, spectral analysis \cite{Binney1984}, Shannon entropy \cite{Shannon:1948,Cao:2024bjk},
recurrence analysis\cite{Eckmann:1987}, 0-1 test\cite{Gottwald:2009} and so on.

The advantage of the Poincar\'{e} map lies in its intuitive reflection of the system's orbital dynamics,
but it is only applicable to systems with fewer than three degrees of freedom.
A single point on the section represents a strict periodic orbit, while a single or multiple closed curves indicate a quasiperiodic system.
If the points are randomly distributed, the system is chaotic. The Lyapunov exponent is an indicator that measures the average rate at which two nearby trajectories diverge
over time, and it reflects the intensity of chaos. In the context of general relativity, the two-particle method is typically used for its calculation, as the variational
method requires the use of the geodesic deviation equation and the derivation of complex curvature tensors.

When the motion of the celestial bodies under study does not necessarily remain within a bounded region, such as in the case of the general three-body problem, where the final
evolution results in the formation of a binary system and a third body that either escapes or moves away, the Lyapunov exponent of the system tends to zero as time approaches
infinity. Thus, discussing the chaos of the system may seem meaningless.
However, by starting from a single orbit and dividing the total integration time into several small intervals, the Lyapunov exponent for each interval can be computed
separately. This allows for the calculation of a series of local Lyapunov exponents, which in turn provides the distribution of the local Lyapunov exponent spectrum.
The Fast Lyapunov Indicator (FLI) is an extension of the Lyapunov exponent in curved spacetime, and renormalization is required for chaotic orbits.
The FLI grows exponentially with time, indicating the chaotic nature of the orbit; when the FLI grows algebraically with time, it signifies the regularity of the orbit.

The frequency spectrum analysis is a method that transforms a time series into the frequency domain and analyzes its characteristics.
It is also an important numerical tool for describing the global dynamic features of multi-dimensional systems.
The calculation of Shannon entropy is similar to that of the local Lyapunov exponent, both relying on constructing probability distributions and performing calculations over
intervals. In principle, this does not require knowledge of the dynamical equations, only time series data.
The entropy fluctuation of chaotic trajectories is significantly stronger than that of order trajectories.
  {In contrast, recurrence analysis effectively captures the repetitive structure of the phase space. Its core involves embedding the time series into the phase
space, calculating the distances between state points, and generating a recurrence plot by setting a threshold: when the distance between two points is smaller than the
threshold, they are marked as recurrence points. Short-term determinism in chaotic systems is manifested by recurrence points arranged along the diagonal, while a long
diagonal suggests the deterministic dynamics of the system.
The 0-1 test provides a binary chaos criterion by calculating the diffusion rate of the time series in an auxiliary space, directly outputting 0 (indicating regular) or 1
(indicating chaotic). It avoids the difficulty of selecting parameters for phase space reconstruction, making it suitable for rapid screening.
Due to the complexity of the geodesic equations in curved spacetime, especially for quasi-periodic motion that is difficult to distinguish from chaos, it is more reliable to
identify it using multiple chaos indicators.}

As a substantial broaden and deepen of our prior study \cite{Cao:2024bjk}, we further introduce the basic concept of mutual information into the calculation of orbital
dynamics to measure the uncertainty of the system. According to the principles of information thermodynamics, before the measurement, the statistical state of the system is
described by the probability distribution $\rho(x)$, and after the measurement, the state of the system changes to the conditional probability distribution $\rho(x|m)$.
Therefore, the change in the system's entropy can be
expressed as
$\Delta S=H(\rho(x|m))-H(\rho(x))$. According to the formula of information thermodynamics \cite{Parrondo:2015, Cai:2017ihd}, the entropy change caused by the measurement is
inversely
proportional to the change in the mutual information,
\begin{eqnarray}
\Delta S=k(H(\rho(X|M))-H(\rho(X)))=-kI(X;M),
\end{eqnarray}
where $k$ represents the Boltzmann constant, and $I(X;M)$ represents the mutual information.
The change in entropy during the measurement is inversely proportional to the mutual information,
reflecting how the acquired information influences the uncertainty of the system.
For the relativistic system studied in this paper, if the orbit is in a chaotic state before and after measurement, it indicates that the mutual information gained from the
measurement is small, and the system's entropy remains largely unchanged, reflecting a high degree of uncertainty in the orbit. On the other hand, if the orbit is in an order
state, the mutual information obtained from the measurement is larger, and the system's entropy will decrease significantly, reflecting the orderliness of the orbit. By using
the particle pair method, it may be possible to determine the orbital state.

In Section \ref{sec:two}, the well-known Wlad potential and Kerr spacetime are introduced. Section \ref{sec:three} briefly discusses how to construct the explicit symplectic
algorithm in curved spacetime and its physical significance. Section \ref{sec:four} provides a detailed explanation of the birth of the MIPP method. In Section \ref{sec:five},
we demonstrate the powerful ability of MIPP in identifying orbital states. Finally, in Section \ref{sec:six}, we conclude with a summary of this method.

\section{Kerr black hole with external magnetic field}\label{sec:two}
The motion of a relativistic charged particle with rest mass $m$ and charge $q$
around the Kerr black hole in the presence of  an external magnetic field with strength $B$ is described by the
following Hamiltonian formalism \cite{Carter:1968ks}
\begin{eqnarray}
H &=& \frac{1}{2m}g^{\mu\nu}(P_{\mu}-qA_{\mu})(P_{\nu}-qA_{\nu}),
\end{eqnarray}
where the contravariant Kerr metric $g^{\mu\nu}$ has nonzero components \cite{Sun:2021oxg,Cao:2022bvu}
\begin{eqnarray}
&&
g^{tt}=\frac{1}{\Sigma}[a^2\sin^{2}\theta-\frac{(r^2+a^2)^2}{\Delta}],
\\
&&
g^{t\phi}=\frac{a}{\Sigma}(1-\frac{r^2+a^2}{\Delta})=g_{\phi
t},
\\
&&
g^{rr}=\frac{\Delta}{\Sigma},
\\
&&
g^{\theta\theta}=\frac{1}{\Sigma},
\\
&&g^{\phi\phi}=\frac{1}{\Sigma}(\frac{1}{\sin^{2}\theta}-\frac{a^2}{\Delta}),
\\
&&\Sigma=r^2+a^2\cos^{2}\theta,
\\
&&\Delta=r^2+a^2-2Mr.
\end{eqnarray}
 Here $A^{\mu}$ is the well-known Wald potential \cite{Wald:1974np} that satisfies the source-free Maxwell equations in vacuum spacetime
\begin{equation}
A^{\mu}=aB\xi^{\mu}_{(t)}+\frac{B}{2}\xi^{\mu}_{(\phi)},
\end{equation}
where $\xi^{\mu}_{(t)}$ and $\xi^{\mu}_{(\phi)}$ are time-like and spacelike Killing vectors, respectively.
In the case of Kerr spacetime, the nonzero covariant components of the four-vector potential are given by\cite{Sun:2021oxg}
\begin{eqnarray}
&&A_{t}=aBg_{tt}+\frac{B}{2}g_{t\phi} \nonumber\\
&&=-aB[1+\frac{r}{\Sigma}(\sin^2\theta-2)] \\
&&A_{\phi}=aBg_{t\phi}+\frac{B}{2}g_{\phi\phi} \nonumber\\
&&=B\sin^2\theta[\frac{r^2+a^2}2+\frac{a^2r}{\Sigma}(\sin^2\theta-2)].
\end{eqnarray}
Since the time-like 4-velocity always satisfies the relation
\begin{eqnarray}
\dot{x}^{\mu}\dot{x}_{\mu}=-1,
\end{eqnarray}
the Hamiltonian (1) for the mass particle is a constant given by
\begin{eqnarray}
H=-\frac{1}{2}.
\end{eqnarray}
Therefore, in order for the Hamiltonian to remain a constant,  the canonical four-momentum $P_{\mu}$ and the kinematical
four-momentum $p_{\mu}$ have the relation
 \begin{eqnarray}
 P_{\mu}=p_{\mu}+qA_{\mu}=g_{\mu\nu}\dot{x}^{\nu}+qA_{\mu},
\end{eqnarray}
where $P_{t}$ and $P_{\phi}$ are two constants of motion, representing the particle's energy $E$
and angular momentum $L$, respectively
 \begin{eqnarray}
 &&
 P_{t}=g_{tt}\dot{t}+g_{t\phi}\dot{\phi}+qA_{t}=-E,
 \\
 &&
 P_{\phi}=g_{t\phi}\dot{t}+g_{\phi\phi}\dot{\phi}+qA_{\phi}=L.
\end{eqnarray}
Now, Eq. (1) is rewritten as

\begin{eqnarray}
 &&
 H=\frac12g^{\mu\nu}(P_\mu-qA_\mu)(P_\nu-qA_\nu) \nonumber\\
 &&
 =F+\frac12\frac\Delta\Sigma P_r^2+\frac12\frac{P_\theta^2}\Sigma,
\end{eqnarray}
where $F$ is a function of $r$ and $\theta$ as follows:
\begin{eqnarray}
 &&
F=\frac12[g^{tt}(E+qA_t)^2+g^{\phi\phi}(L-qA_\phi)^2]\nonumber\\
 &&
-g^{t\phi}(E+qA_t)(L-qA_\phi).
\end{eqnarray}
Throughout this paper, we adopt the speed of light $c$ and the gravitational constant
$G$ as geometric units, i.e., $c=G=1.$
The black hole mass $M$ and the particle mass $m$  are also set to unity, i.e., $M=m=1$.
The magnetic field parameter $b=qB$.
However, the inclusion of an electromagnetic field into the system breaks the usual conservation of integrals of motion,
such as the Carter constant, which is essential for the system's integrability.
As a result, the system enters a non-integrable regime, leading to the onset of chaotic dynamics.
The emergence of chaos significantly complicates the analytical prediction of the long-term evolution of particle trajectories. Therefore, to accurately capture the system's
dynamics,
it is necessary to employ reliable numerical methods to simulate and analyze the behavior of the system.
\section{Numerical integration scheme}\label{sec:three}
Based on the work of Wu et al. \cite{Wang:2021gja, Wang:2021xww, Wang:2021yqk, Wu:2021rrd, Wu:2022nye, Zhou:2022uht, Wu:2024ehd},
we adopt an optimized fourth-order partitioned Runge-Kutta (PRK) symplectic algorithm \cite{Zhou:2022uht} to solve the Hamiltonian (20).
This method requires that each sub-Hamiltonian in Eq. (20) be analytically solvable.
The first term $F$ is solved analytically, but the second  and third term are not.
Using the time-transformation function
\begin{eqnarray}
d\tau=T(r,\theta)dw, \;\;T(r,\theta)=\frac{\Sigma}{r^{2}},
\end{eqnarray}
the difficulty can be effectively addressed \cite{Wu:2021rrd}. The time-transformed Hamiltonian is
\begin{eqnarray}
&&
K=T(H+p_{0})=K_{1}+K_{2}+K_{3}+K_{4}+K_{5} \nonumber\\
&&
=\frac{\Sigma}{r^{2}}(F+p_{0})+\frac{1}{2}p_r^2-\frac{1}{r}p_r^2+\frac{a^{2}}{2r^{2}}p_r^2+\frac{1}{2r^{2}}p_\theta^2=0,
\end{eqnarray}
$p_{0}=-H$ is a momentum with respect to  coordiante $\tau$. With these analytically sub-Hamiltonians,
an explicit symplectic integrator can be obtained. From the above simple analysis, we can conclude that the time-transformation function $T(r,\theta)$ can be adjusted
according to the specific needs of the physical model and is not strictly constrained to be equal to 1. Specifically, when the Hamiltonian is zero, the time-transformation
function operates on the Hamiltonian in a manner analogous to a conformal transformation of the metric $g_{\mu\nu}$. That is:
\begin{eqnarray}
&&
0=K=\Omega^{2}(H+p_{0})\nonumber
\\
&&
=\Omega^{2}[\frac12g^{\mu\nu}p_\mu p_\nu+\frac12g^{\tau\tau}p_{\tau}p_{\tau}]\nonumber
\\
&&
=\frac12\Omega^{2}g^{mn}p_m p_n\nonumber
\\
&&
=\frac12\tilde{g}^{mn}p_m p_n, \;\; m,n=\tau,t,r,\theta,\phi
\end{eqnarray}
it's worth noting that the conformal transformation factor $\Omega^{2}$ must be greater than zero,
which implies that the time-transformation factor $T(r,\theta)$  is also a positive function everywhere.

Therefore, an alternative choice of conformal transformation factor $\Omega^{2}$ is
\begin{eqnarray}
\Omega^{2}=2\Sigma.
\end{eqnarray}
Now, the  time-transformed Hamiltonian  is given by
\begin{eqnarray}
&&
K=K_{1}+K_{2}+K_{3}+K_{4}\nonumber\\
&&
={2\Sigma}(F+p_{0})+r^{2}p_r^2-2rp_r^2+(a^{2}p_r^2+p_\theta^2)=0,
\end{eqnarray}
where the four components of the Hamiltonian are solvable and can also be applied to the $PRK_{64}$ algorithm.
We then evaluated the energy error performance of the algorithm for particles released from different initial positions.
As shown in Fig. 1(a), when the step size $h$ is chosen as $0.3/\Omega^{2}$,
the energy error for order orbits(Orbit 1) remains stable at approximately $10^{-9}$,
demonstrating the algorithm's high precision in regular motion.
For chaotic orbits(Orbit 2), the energy error remains consistently low  and exhibits negligible differences across the three different step sizes as shown in Fig. 1(b).
Thus, we selected the algorithm with a step size of $0.3/\Omega^{2}$ to investigate the chaotic dynamics in Kerr spacetime.

  {In Figures 1 and 2, under the guarantee of high-precision algorithms, we subsequently computed the Poincar\'{e} section of the orbit,
Shannon entropy along with its spectral analysis, as well as the time series data of the orbit and its power spectrum.
The fluctuations of Shannon entropy for Orbit 1 are small, and the closed curve on the Poincar\'{e} section indicates that the orbit is in an order state.
In contrast, Orbit 2 exhibits significant fluctuations in Shannon entropy and random scattering points on the section, indicating its chaotic nature.
It is not possible to accurately determine the true state of the orbit by merely observing the time series data in Figure 2.
The power spectrum of the time series clearly reveals the results: in Fig. 2(c), the presence of a few frequency components indicates that Orbit 1 exhibits periodic motion;
whereas in Fig. 2(d), the increase in frequency components, the denser spectral lines, and the appearance of distinct continuous spectra suggest that Orbit 2 is a chaotic
trajectory.
Similar conclusions can be drawn by examining the entropy spectra in Fig. 2(d) and 2(f).
Although the above chaotic indicator methods are reliable, they all depend on visual inspection,
which becomes cumbersome when analyzing the motion states of a large number of particles.
This indicates that, in addition to requiring high-precision numerical algorithms, we also need fast chaotic indicators to quickly identify a large number of orbits.}
\section{Mutual Information for Particle Pair} \label{sec:four}
In this section, we will provide a detailed description of the implementation of the MIPP (Mutual Information for Particle Pair) and explain its application for orbits around
Kerr black holes in the next section.
\subsection{The use of Shannon entropy}
To better understand the MIPP method, we begin by introducing Shannon entropy as a tool for detecting chaotic phenomena in curved spacetime \cite{Cao:2024bjk}.
In general, data or time series obtained from physical systems will fall within a specific range.
By dividing this range into small intervals, we can construct a probability distribution based on how the data points are distributed across these intervals, making the
calculation of Shannon entropy straightforward.

\subsection{Two-Particle method}
The two-particle method is a numerical technique used to study the trajectory dynamics of complex dynamical systems\cite{Tancredi:2001},
and it has been widely applied in various settings, including curved spacetimes \cite{Wu:2006rx}.
The core idea of this method is to simultaneously track the trajectories of two nearby particles
over time. Compared to traditional variational methods used for chaotic systems, the two-particle method offers significant advantages in computing the maximum Lyapunov
exponent. Variational methods require deriving the system's variational equations, which often involve intricate mathematical derivations. In contrast, the two-particle method
directly tracks the trajectories of two nearly identical particles with slightly perturbed initial conditions and calculates their separation distance, thereby simplifying the
computational process and reducing its complexity.

\subsection{The calculation of Mutual Information}
The mutual information (MI) is an important concept in information theory, used to measure the correlation or dependency between two random variables.
Given two discrete random variables $X$ and $Y$, with joint distribution $\rho(x,y)$ and  marginal distributions
$\rho_{x}$ and $\rho_{y}$, the mutual information $I(X;Y)$ is defined as \cite{Parrondo:2015}:

\begin{eqnarray}
&&
I(X;Y)=\sum_{x\in X}\sum_{y\in Y}\rho(x,y)\log\left(\frac{\rho(x,y)}{\rho(x)\rho(y)}\right) \nonumber \\
&&=H(X)+H(Y)-H(X,Y),
\\
&&
\bar{I}(X;Y)=\frac{I(X;Y)}{H(X,Y)},
\end{eqnarray}
where $H(X)$ and $H(Y)$ represent the entropies of the random variables $X$ and $Y$, respectively,
$H(X,Y)$ is the joint entropy, which quantifies the uncertainty of their joint distribution,
and $\bar{I}(X;Y)$ represents the normalization to $I(X;Y)$.
It is always non-negative, symmetric, and vanishes if an only if $X$ and $Y$ are statistically independent.
  {In chaotic systems, due to exponential sensitivity to initial conditions, small perturbations in the initial state cause the trajectories to diverge at a rate
of $e^{\lambda t}$. As a result, in the long-term limit, the mutual dependence between variables tends to vanish. Consequently, the joint distribution of the system approaches
the product of the marginal distributions of individual variables, leading to the mutual information
$I(X;Y)$ approaching zero. indicating that the statistical dependence between variables is completely broken. In contrast, in ordered systems, due to the deterministic
behavior of the system, the trajectories exhibit strong correlations. In this case, the joint entropy H(X,Y) is equal to $H(X)$(or $H(Y)$)
, and the mutual information reaches its theoretical maximum, reflecting the system's complete predictability.}
Thus, we develop the two-particle method to MIPP by using  orbital Shannon \cite{Cao:2024bjk}.
Throughout this paper, the initial separation of $r$ is set to $10^{-8}$,
although other values have been tested, as shown in Fig. 3.
The appropriate choice of initial separation is based on the fact that for order orbits, the MIPP calculation results are close to one, while for chaotic orbits, they are
close to zero.

\section{A Comparative Study of MIPP and FLI }\label{sec:five}
FLI is the generalization of the Lyapunov exponent in curved spacetime.
It is capable of accurately identifying chaotic orbits has been extensively validated in numerous studies
\cite{Chen:2016tmr, Li:2018wtz, Zhang:2023lrt, Cao:2024ihv, Hu:2021gwd, Yi:2020shw, Sun:2021oxg, Sun:2021ndd}. For chaotic orbits, the FLI increases exponentially with time,
whereas for order orbits, the FLI increases algebraically with time.
As a new chaotic indicator, MIPP is used to identify chaotic phenomena, based on the principle that knowing the information of one trajectory can reduce the uncertainty of
another trajectory.
In other words, MIPP measures the degree of correlation between two trajectories.
For chaotic trajectories, the MIPP value tends to zero, while for order trajectories, the MIPP value tends to one.
To compare and analyze the two indicators, MIPP and FLI, we scan the parameter space of the Kerr black hole, using the results obtained from FLI as the reference standard to
verify whether the MIPP results are consistent with those from FLI.

Fig. 4 shows that as the initial release position $r$ of the particle increases,
the orbital state transitions from order to chaotic.
For FLI, the boundary value between chaotic and order states has been experimentally verified as 10.
The scan results indicate that MIPP and FLI align well, especially for some specific points.
For example, at $r$=25, the actual value of FLI exceeds 100, and the MIPP approaches zero,
reflecting the chaotic nature of the orbit.
Another noteworthy point is at $r$=65, where the actual value of FLI is close to, but slightly less than, 10,
while MIPP approaches 1, indicating the order nature of the orbit.

It is worth noting that when determining the transition of the orbit's state, conclusions cannot be drawn solely from the FLI scan plot, as the FLI in the scan is obtained
through the final numerical integration. In such cases, the complete FLI must be output to assess the orbit's state. Therefore, the process of using FLI to numerically
determine the boundary between periodic and chaotic orbits is quite time-consuming.
However, MIPP does not have this issue, it can determine when the transition in the orbit's state occurs based solely on the scan plot.

Fig. 5 shows that as the energy $E$ increases, the chaotic behavior of the particle motion becomes more pronounced, with the results of MIPP and FLI being completely
consistent.   To show the universality of MIPP method, we explore the fundamental case, i.e., the MIPP for orbits in Schwarzschild spacetime.
The result is displayed in Fig. 6. The scan plots  demonstrate a similar trend, further corroborating the universal relationship between energy and chaotic dynamics.
When $E>0.998$, the MIPP value approaches 0.5 compared to 0, but remains less than 0.5. The MIPP for order orbits is always 1.
We expect that when the MIPP result is slightly below or above 0.5, it may indicate a failure of MIPP, and other methods (such as  Poincar\'{e} map, Shannon entropy, or Lyapunov exponents) should be used to verify the true state of the orbit.

Fig. 7 shows the complex changes in the orbital motion as the particle's angular momentum increases. The complexity of the orbit arises from the variation in the particle's
angular velocity and the effects of gravitational dragging. In this case, the two indicators still exhibit a good level of consistency.
In Fig. 8, as the black hole's spin parameter increases, the effect of gravitational dragging becomes more significant, while the influence of the external electromagnetic
field is weakened. The results of the scan indicate that the particle's motion transitions from chaotic to order states.
\section{Conclusions}\label{sec:six}
In this study, we propose a novel chaotic identification indicator, MIPP, based on the cross-fusion of the two-particle method and mutual information approach, which
demonstrates strong capability in identifying orbital states.
We tested the performance of FLI and MIPP in orbital state identification across different parameter spaces in a static spacetime.
The results revealed a high degree of consistency between the two methods.
Specifically, MIPP can accurately indicate transitions in orbital states, while FLI requires careful judgment of the final state of critical orbits.
In terms of computational time, FLI is slightly slower than MIPP.
The main advantage of FLI lies in its use of the two-particle method, which avoids the need to compute complex variational equations, making it highly versatile.
However, for chaotic orbits, FLI requires periodic renormalization to ensure that the distance between the two particles remains sufficiently small, adhering to the observer
theory in general relativity \cite{Liang:2023}.
In contrast, MIPP retains the advantages of FLI while eliminating the need for renormalization, significantly improving computational efficiency and ease of use.

In classical physics, thermodynamics is rather ineffective for few-body system. In a previous related work, we define Shannon entropy for a single orbit. In the present work
we further define mutual information (entropy) for a particle pair, which is demonstrated to be a sensitive and convenient numerical probe for chaotic motion.  We find that the sensitivity of MIPP is
comparable or higher than the traditional indicators for chaos. And at the same time MIPP can save the computing power compared to previous methods.   {We expect
the application potential of MIPP to cover a wide range of fields, from fundamental physics to astronomical detection. For example, under modified gravitational theory
frameworks, MIPP can analyze the behavior of particles in chaotic dynamics and quantify the impact of different gravitational theories on orbital stability, thereby providing
new dynamical criteria to distinguish between general relativity and alternative gravitational models. In gravitational wave astronomy, MIPP can efficiently identify chaotic
phase transitions or resonance phenomena in the orbital evolution of supermassive black holes and compact objects by tracking the mutual information entropy of particle pairs
in extreme mass ratio inspiral (EMRI) systems, thus providing important dynamical constraints for gravitational-wave wave-form modeling.}
\section*{Acknowledgments}
We would like to thank Dr. Shiyang Hu for numerous helpful discussions.
This work is supported by the National Natural Science
Foundation of China (NSFC) under Grant nos. 12235019, 12275106
and Shandong Provincial Natural Science Foundation under grant No. ZR2024QA032.

\begin{figure*}[htbp]
\center{
\includegraphics[scale=0.25]{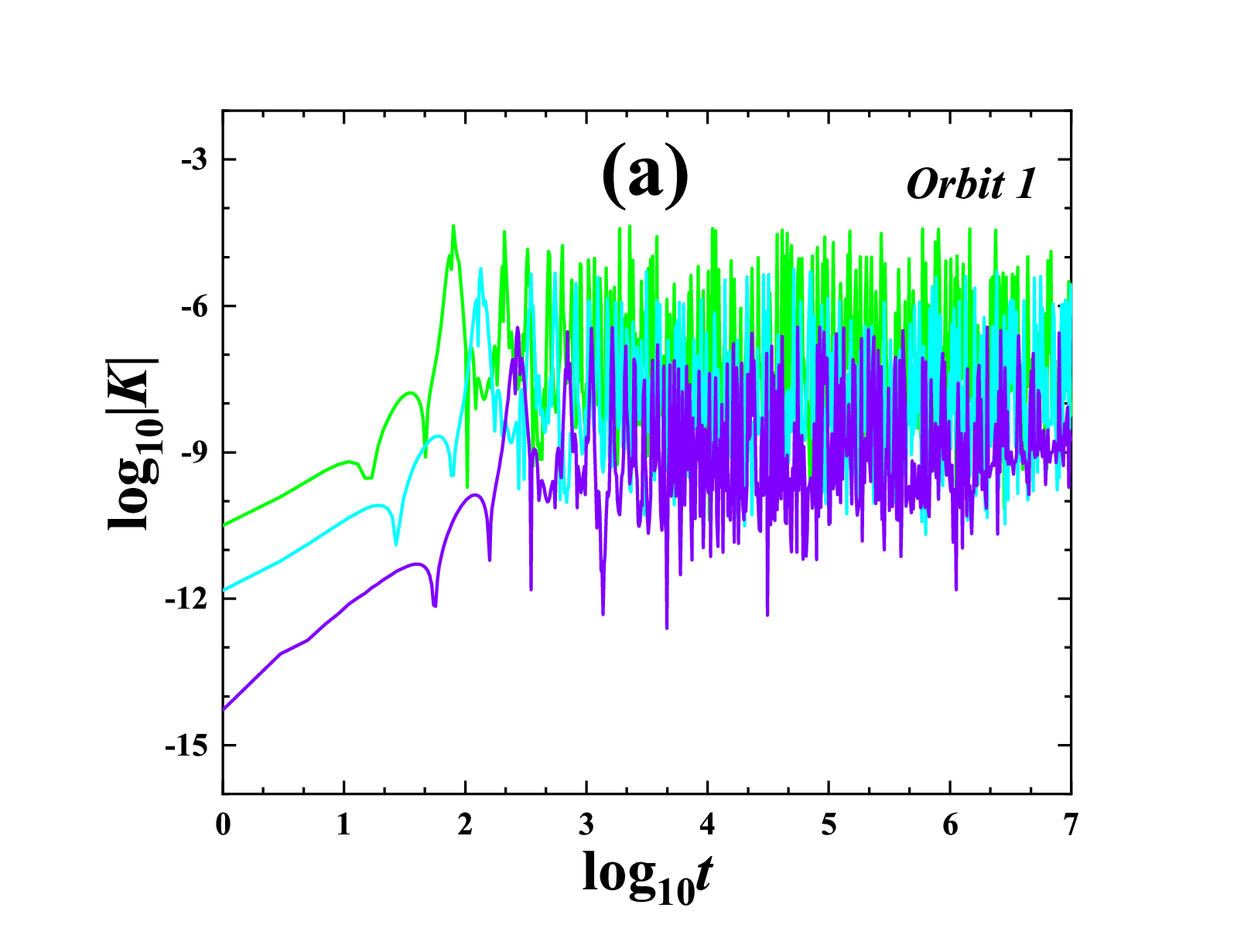}
\includegraphics[scale=0.25]{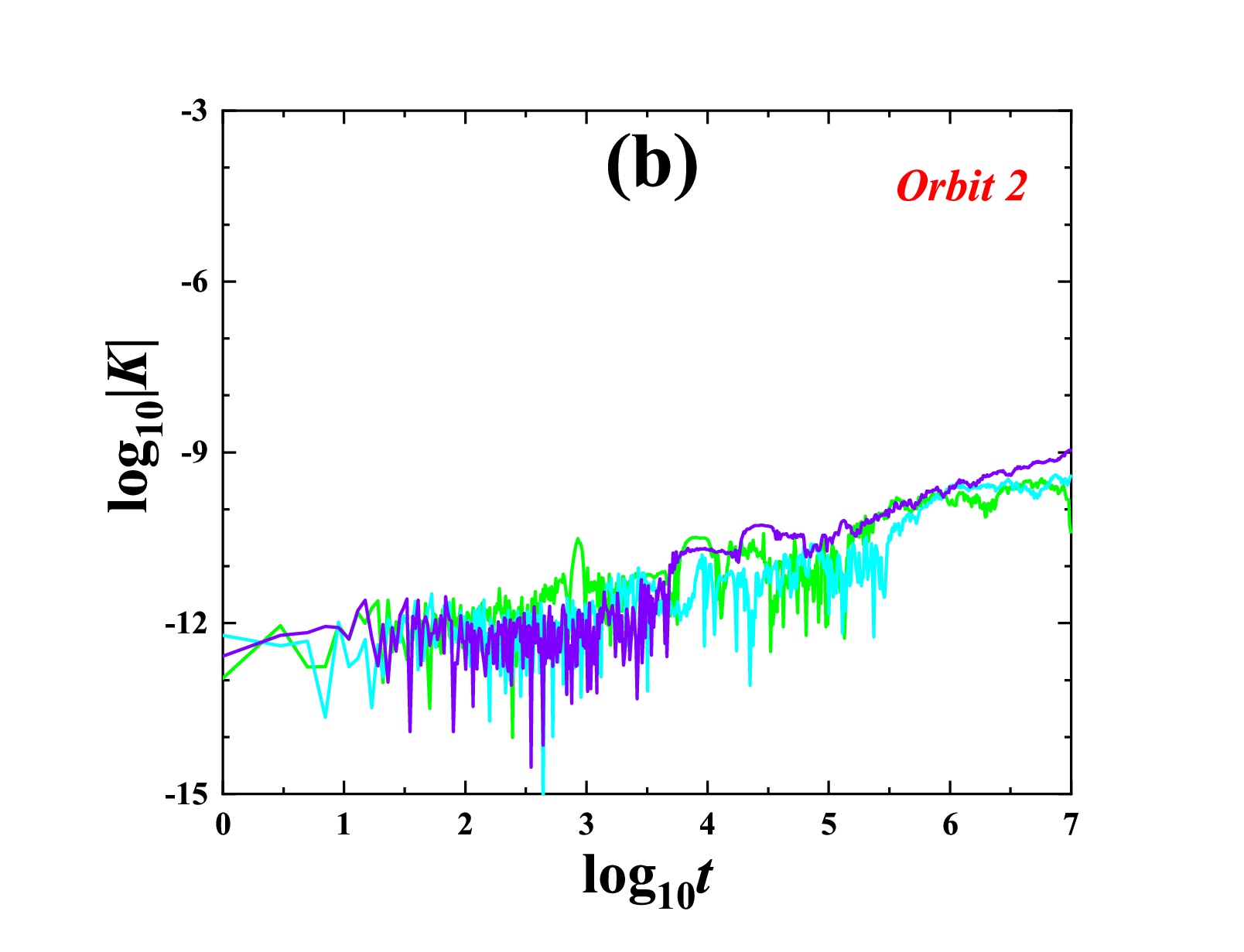}
\includegraphics[scale=0.25]{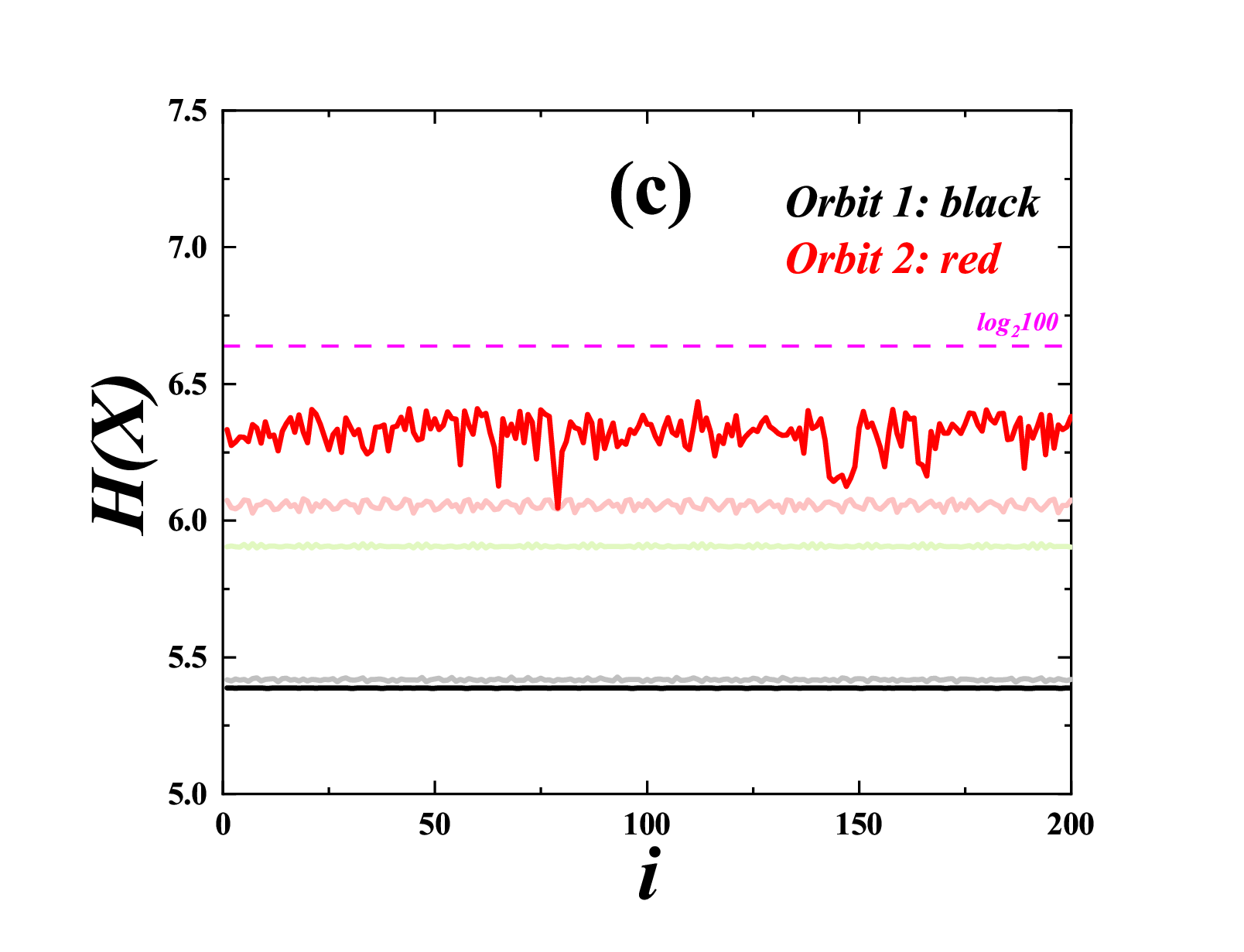}
\includegraphics[scale=0.25]{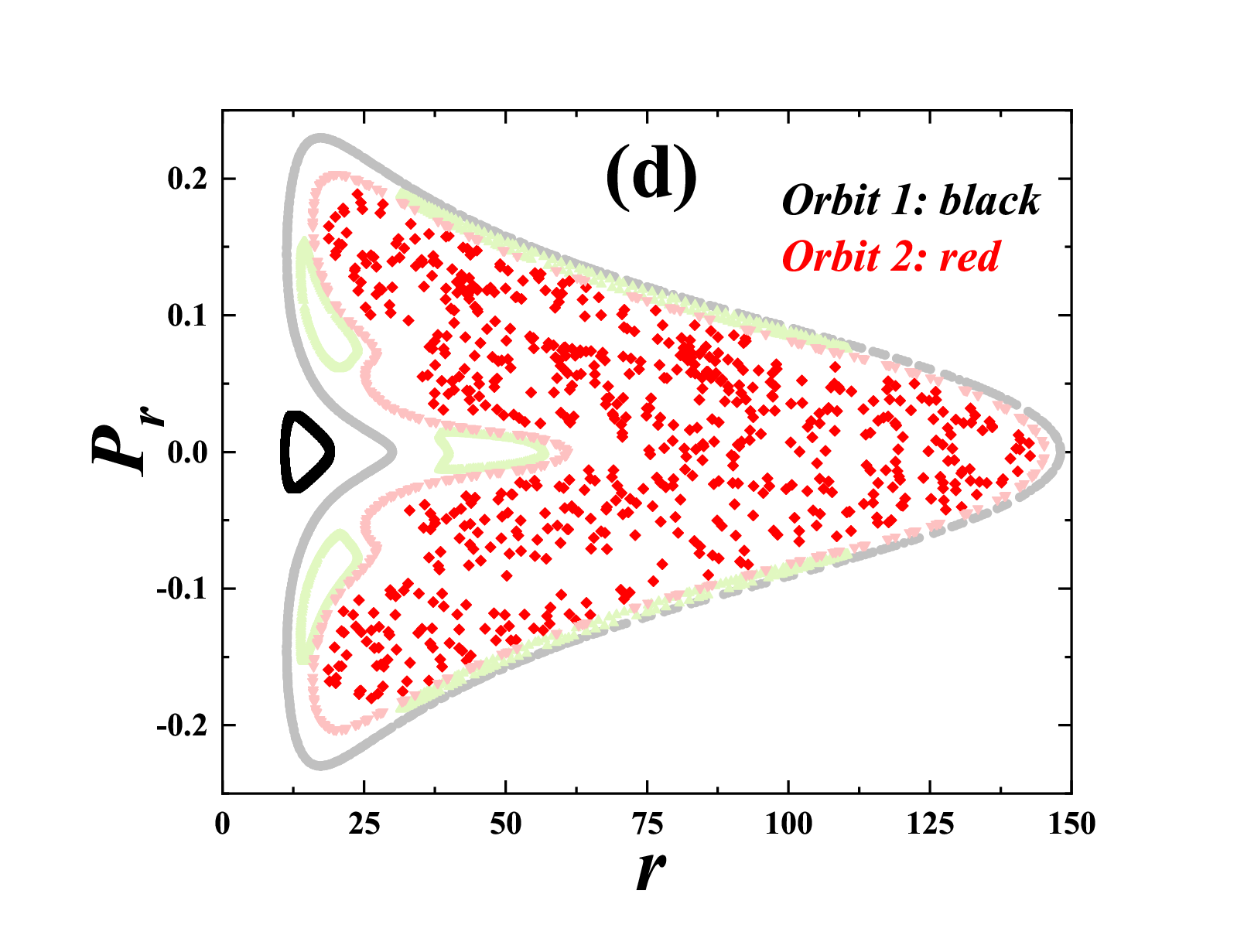}
\caption{Energy errors and chaos indicators for different orbits.
(a) and (b) The orbital parameters are: $E=0.995$, $L=4.6$, $a=0.5$, $b=0.001$, and $\theta=\frac{\pi}{2}$.
Orbit 1 is released at $r=11$, and the Orbit 2 at $r=75$.
The green, blue, and purple curves correspond to step sizes of $1/\Omega^{2}$, $0.6/\Omega^{2}$, and $0.3/\Omega^{2}$, respectively.
(c) and (d) The Poincar\'{e} map and Shannon entropy of the orbits.}
 \label{Fig1}}
\end{figure*}

\begin{figure*}[htbp]
\center{
\includegraphics[scale=0.25]{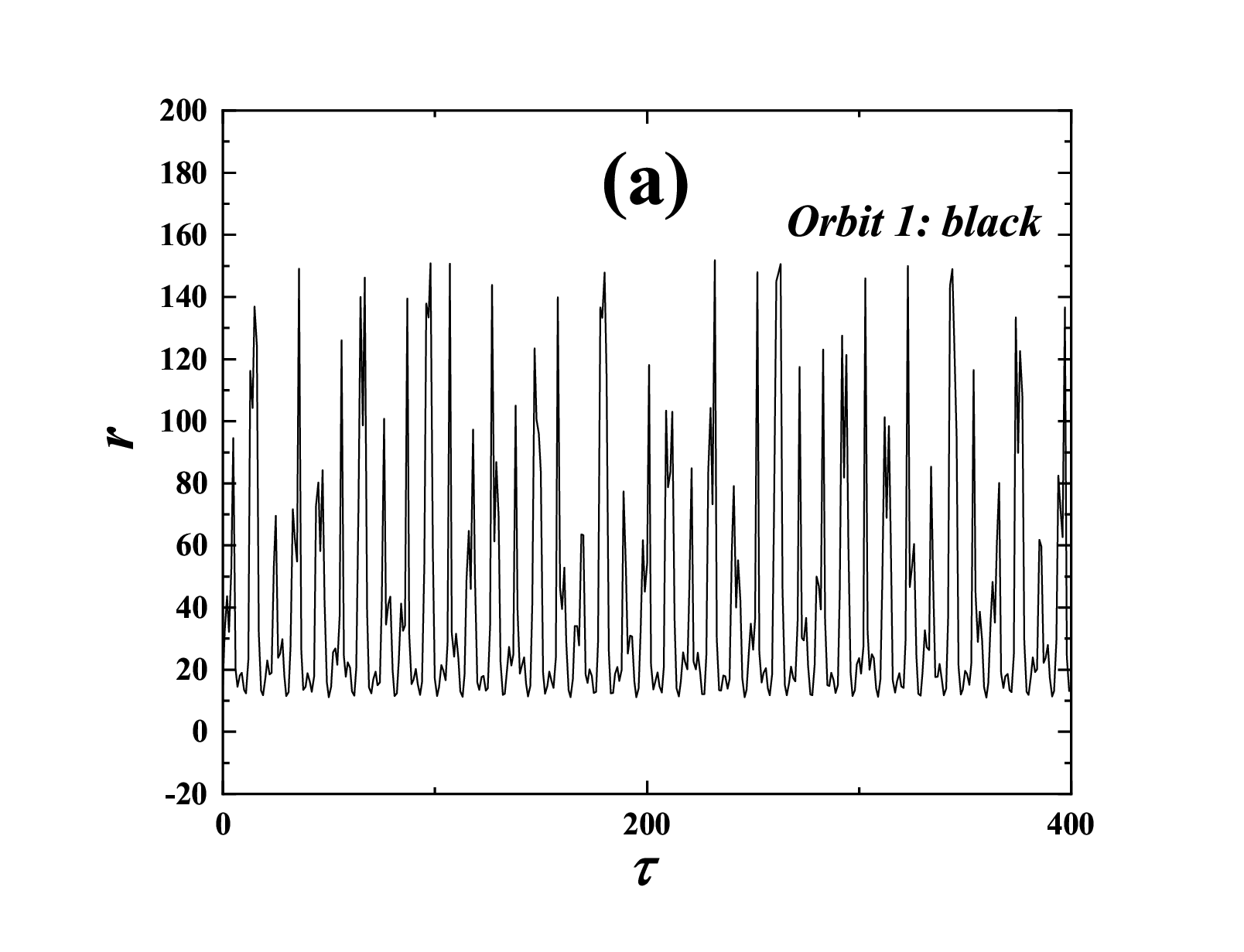}
\includegraphics[scale=0.25]{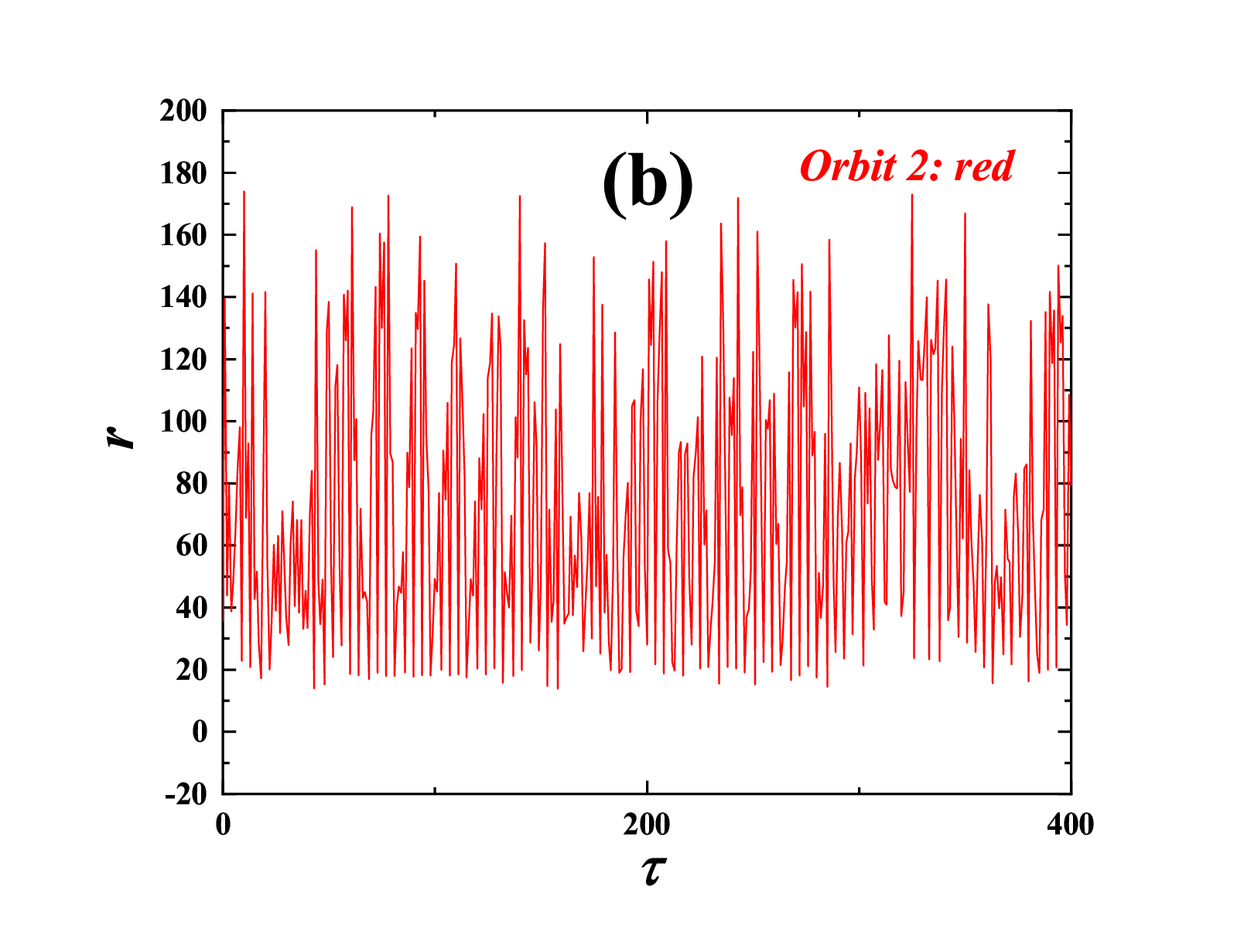}
\includegraphics[scale=0.25]{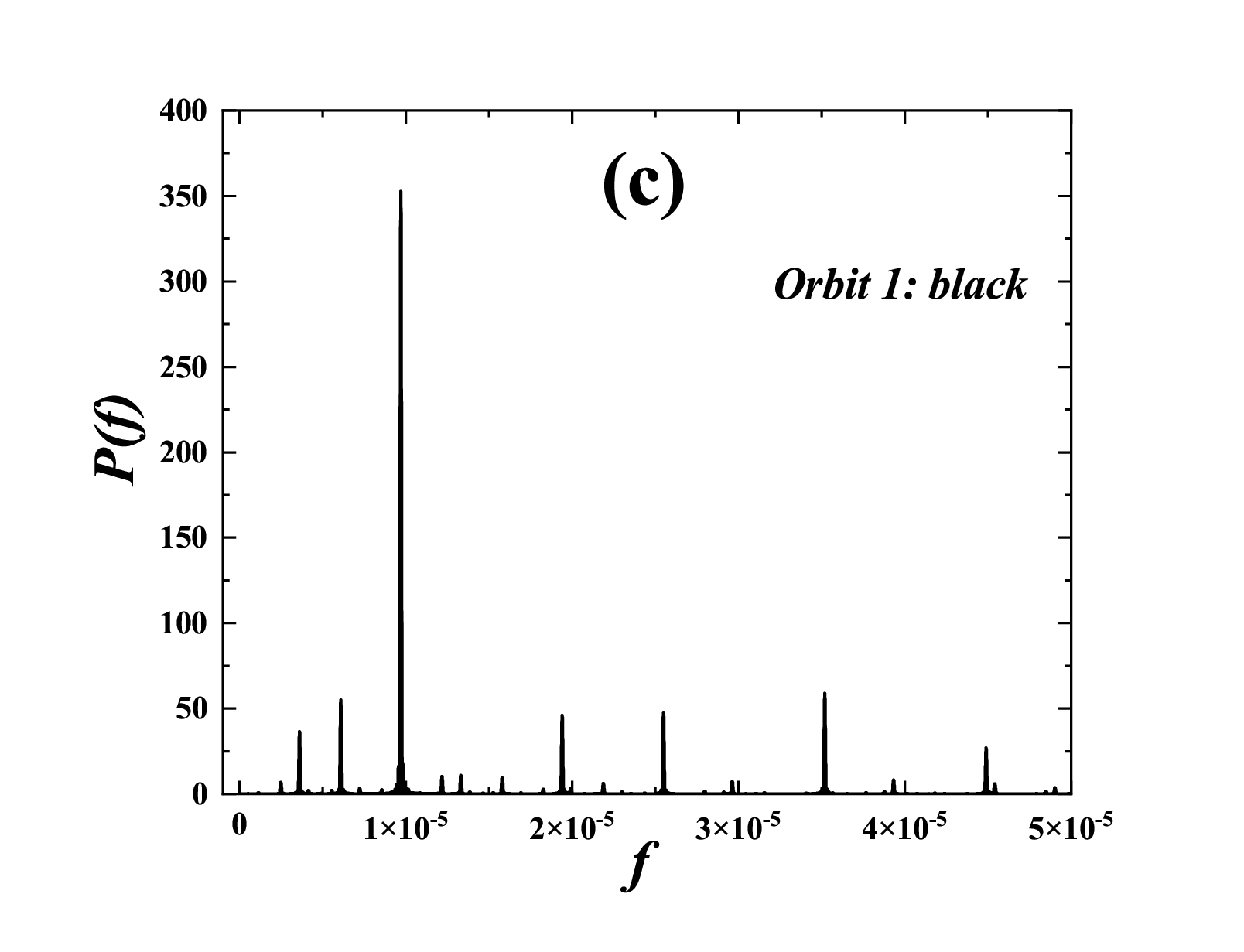}
\includegraphics[scale=0.25]{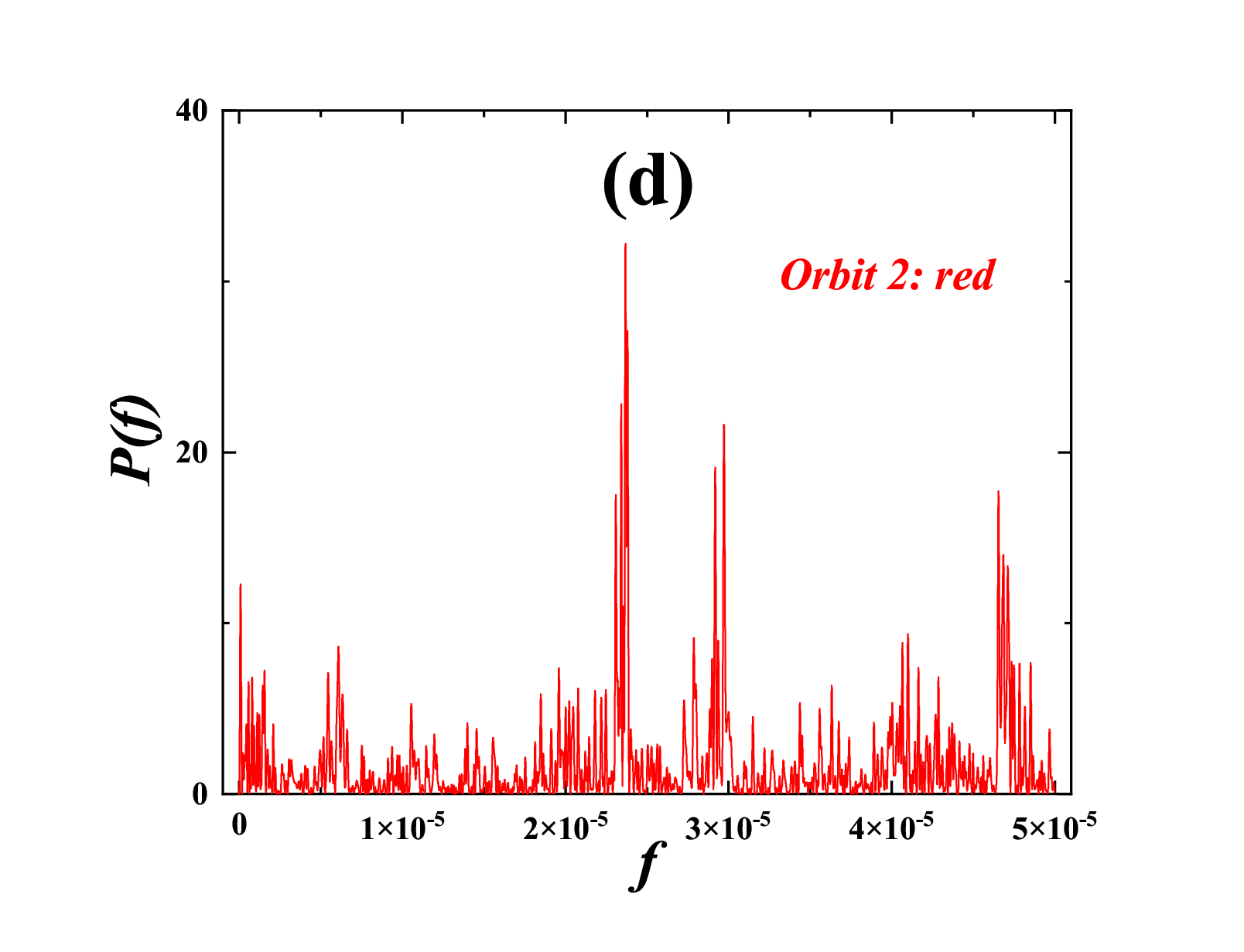}
\includegraphics[scale=0.25]{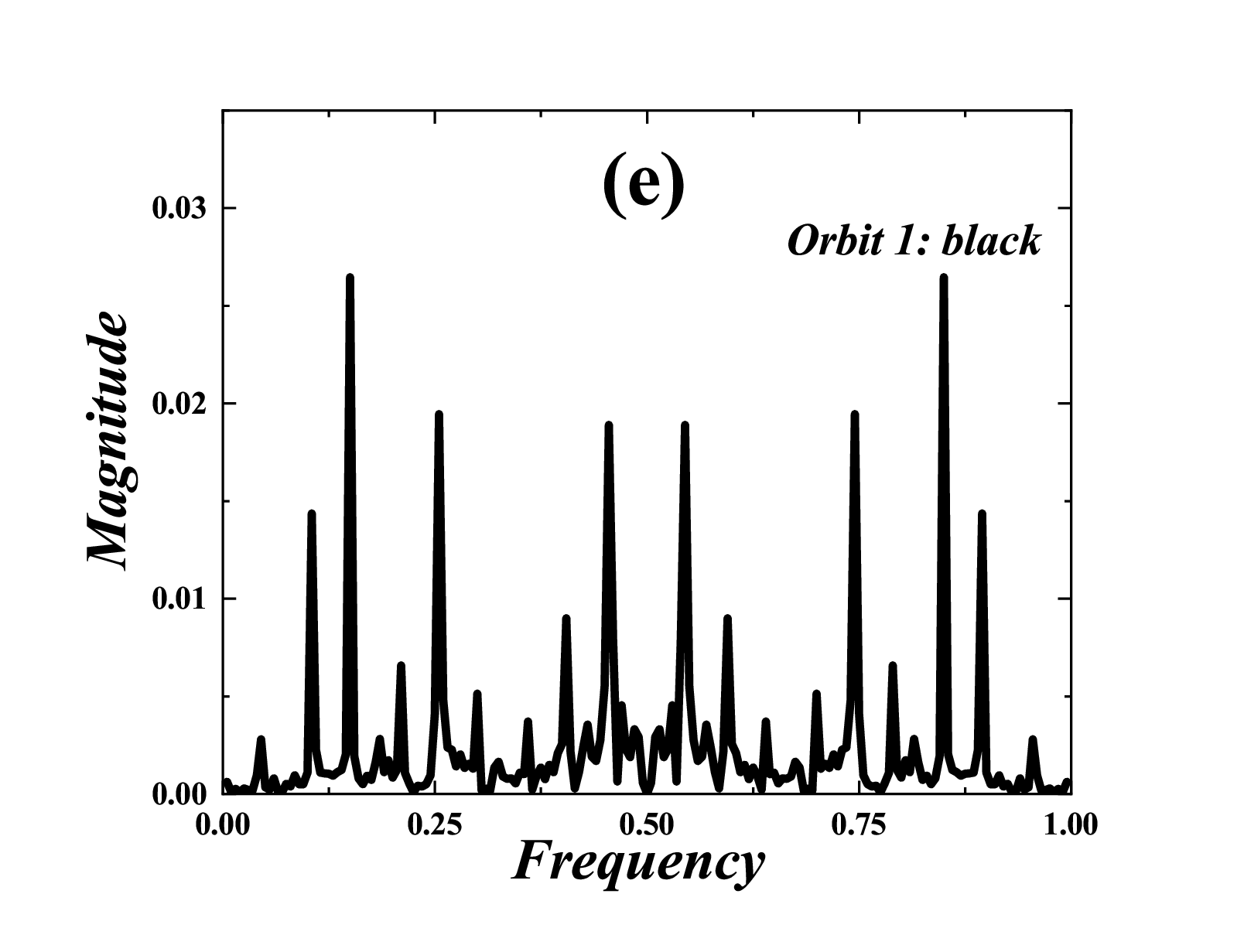}
\includegraphics[scale=0.25]{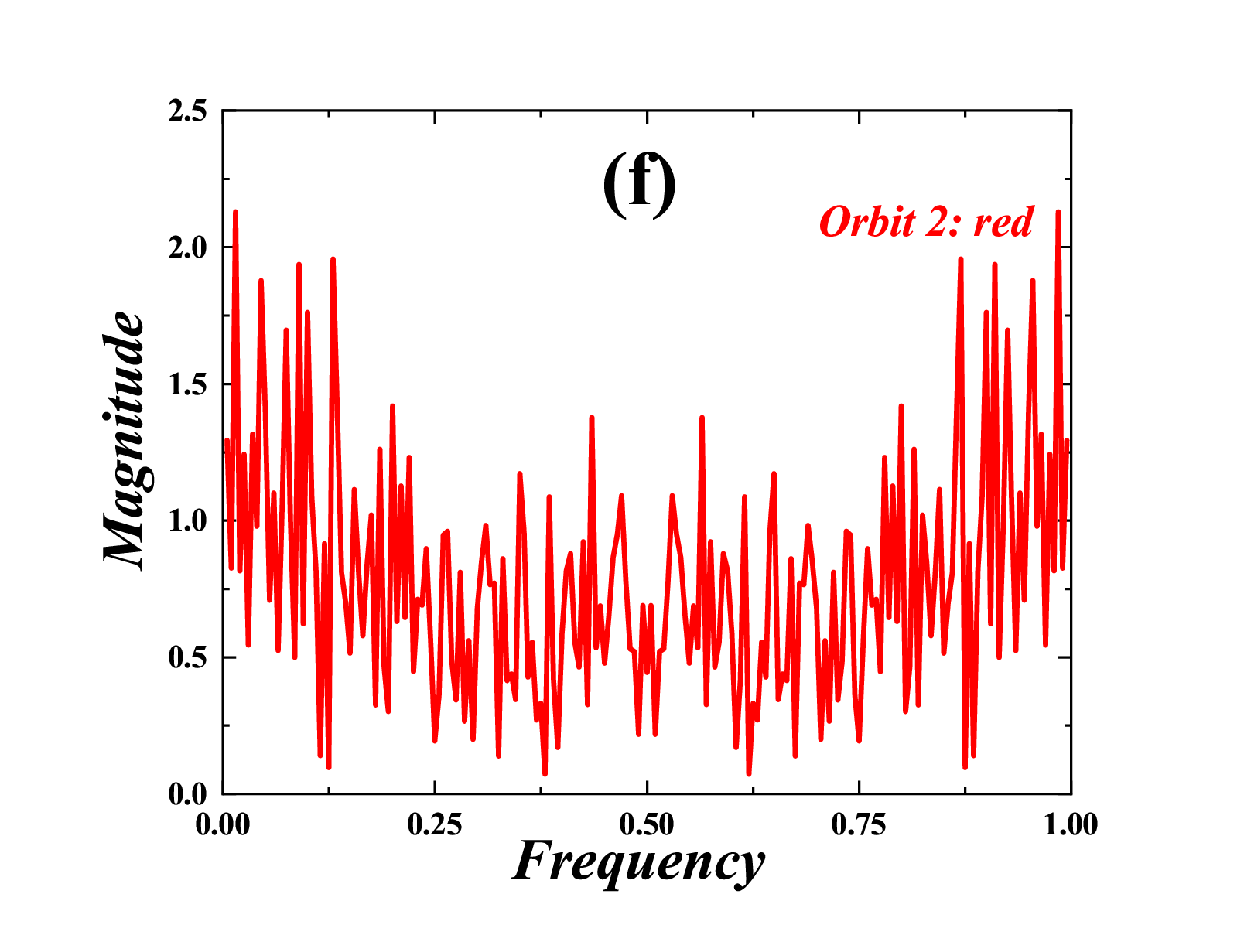}
  {\caption{Time series data and chaos indicators from distinct orbits in Figure (1).
(a) and (b) The variation of r($\tau$) with respect to the proper time $\tau$ for Orbits 1 and 2.
(c) and (d) The power spectrum of the time series.
(e) and (f) The frequency spectrum of Shannon entropy.}
 \label{Fig2}}
 }
\end{figure*}

\begin{figure*}[htbp]
\center{
\includegraphics[scale=0.25]{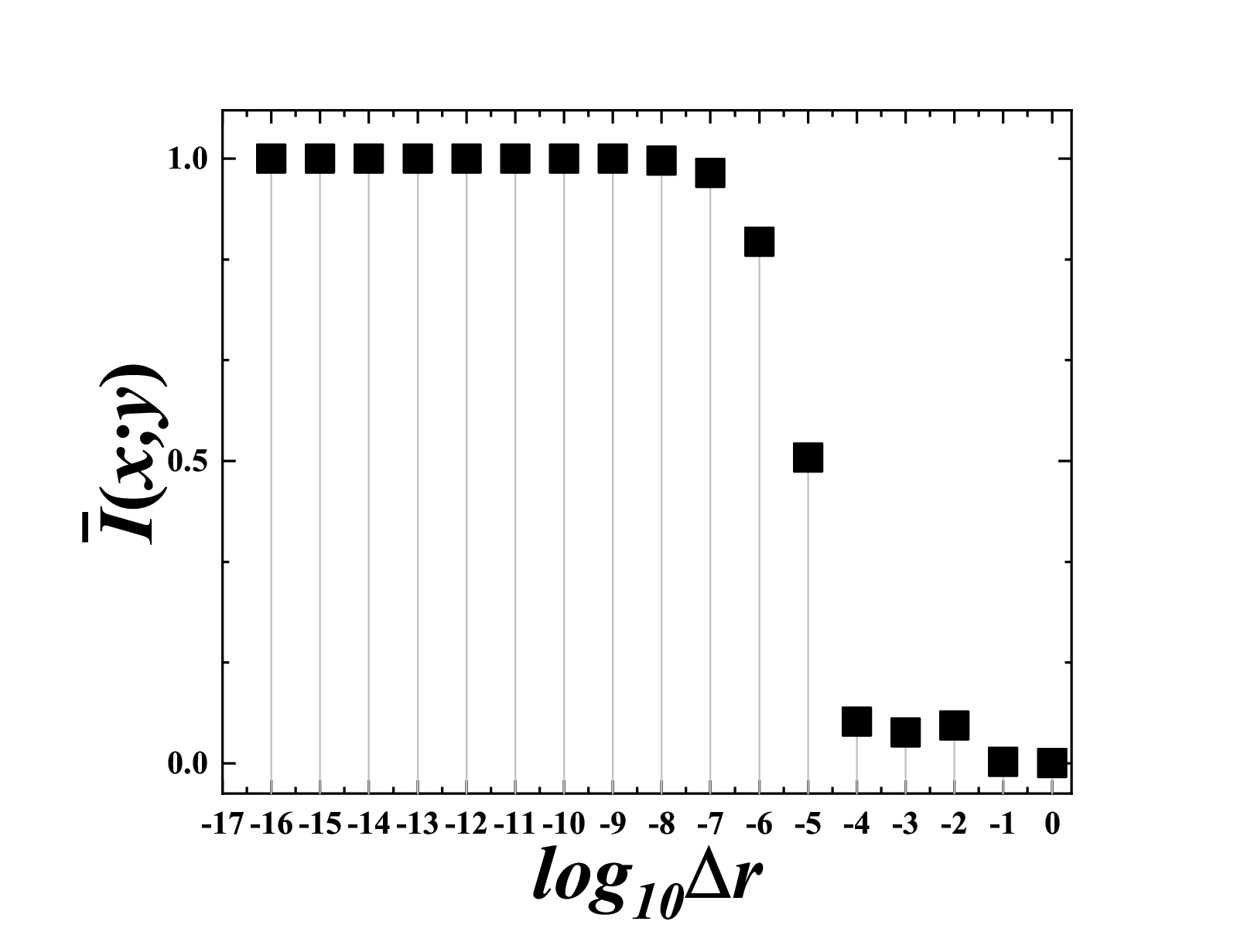}
\includegraphics[scale=0.25]{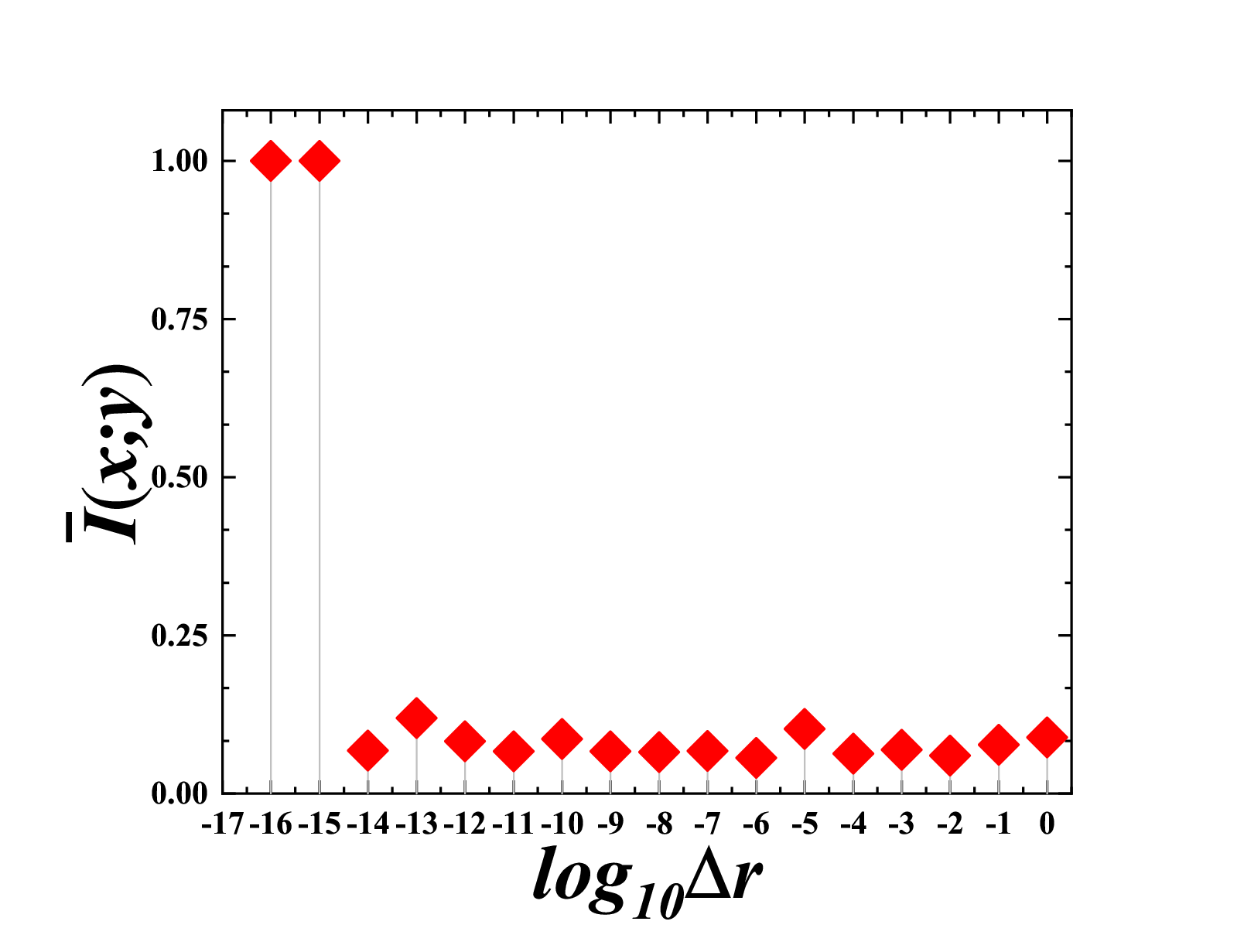}
\caption{From left to right: MIPP of order orbit (Orbit 1) and chaotic orbit (Orbit 2) under different initial separations of $r$. The other orbital parameters are: $E=0.995$,
$L=4.6$, $a=0.5$, $b=0.001$, and $\theta=\frac{\pi}{2}$. }
 \label{Fig3}}
\end{figure*}

\begin{figure*}[htbp]
\center{
\includegraphics[scale=0.25]{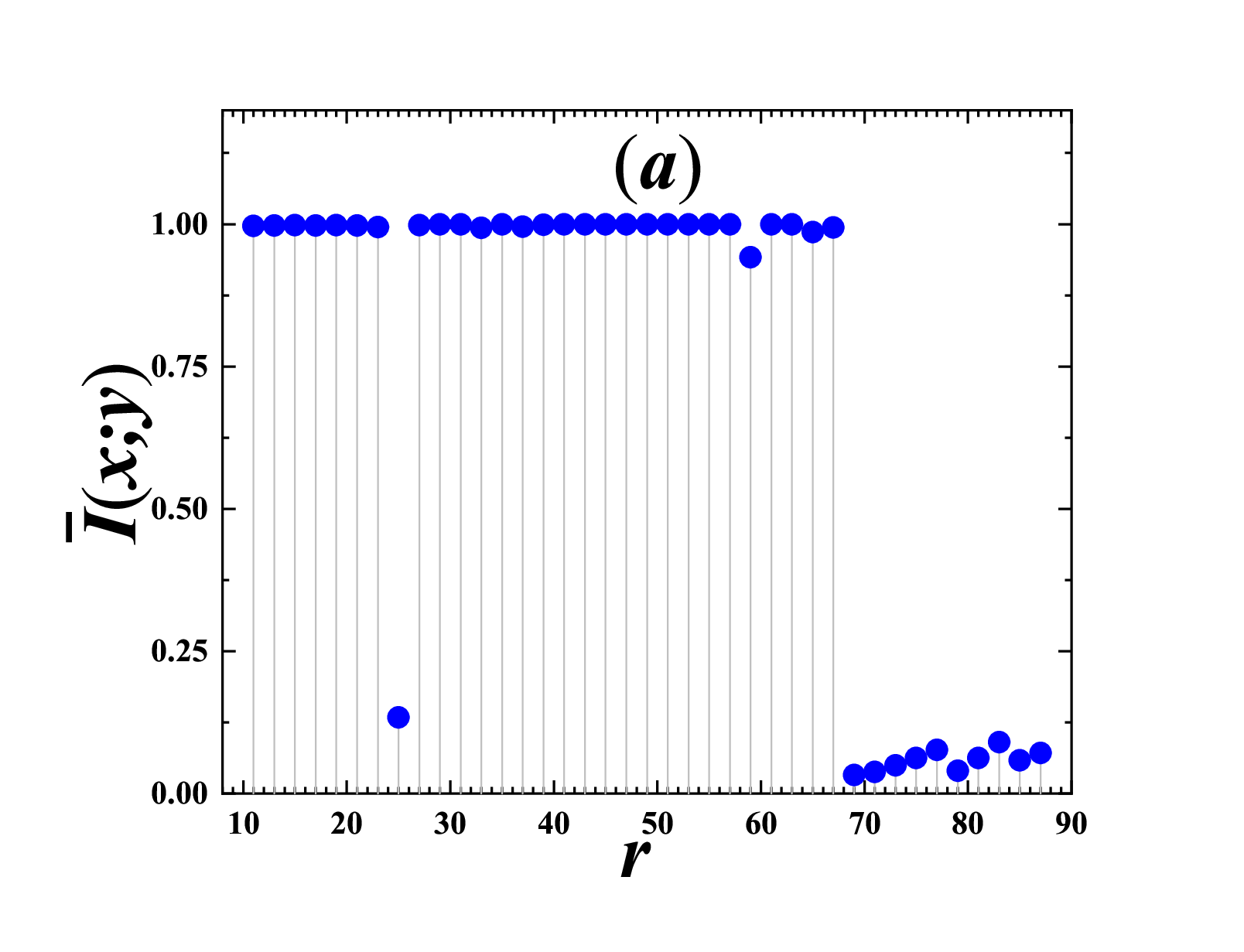}
\includegraphics[scale=0.25]{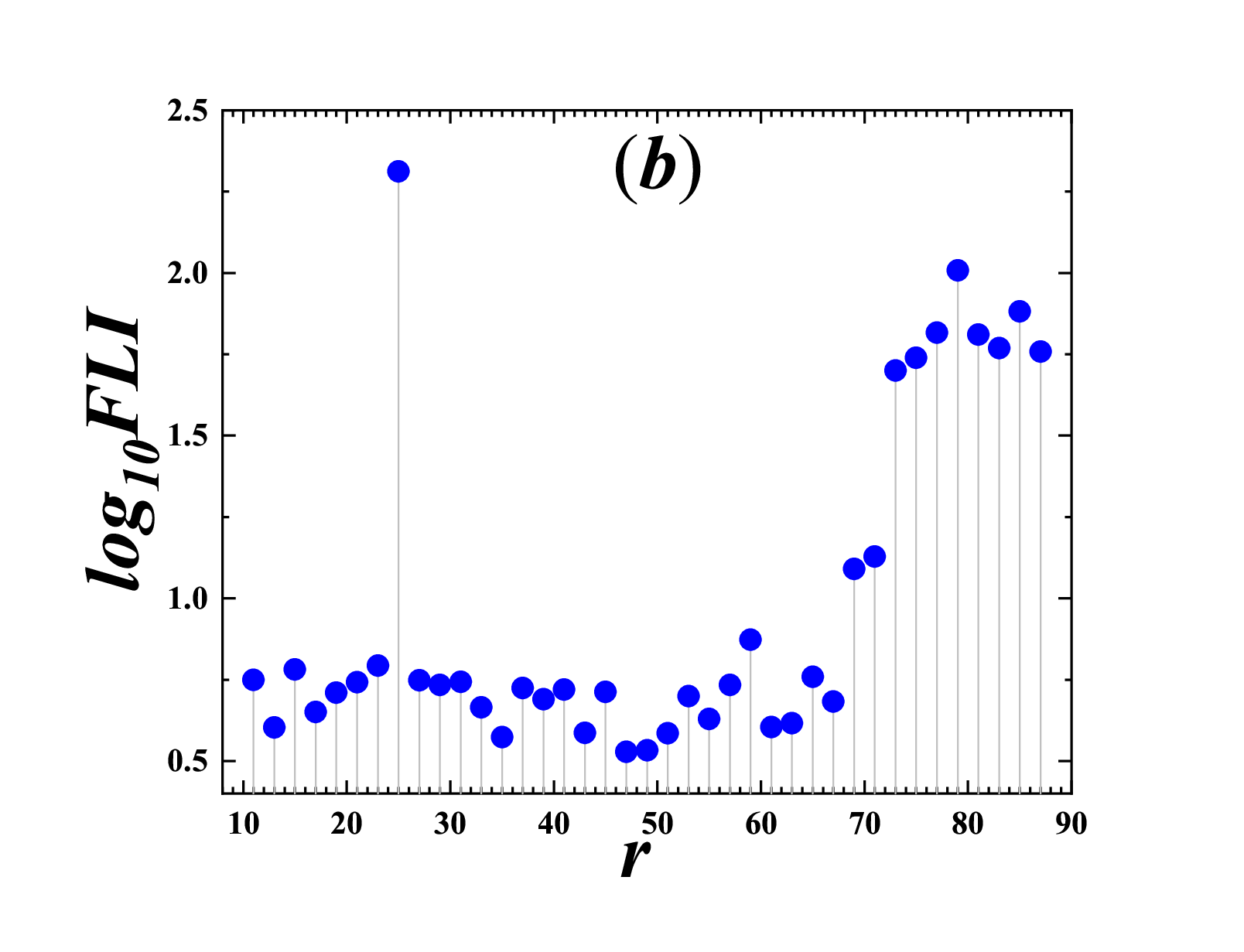}
\caption{MIPP(Cpu cost: 736s) and FLI(Cpu cost: 740s) scan plots for the particle initial release position $r$. The other orbital parameters are: $E=0.995$, $L=4.6$, $a=0.5$,
$b=0.001$, and $\theta=\frac{\pi}{2}$. A total of 39 orbits were studied.}
 \label{Fig4}}
\end{figure*}

\begin{figure*}[htbp]
\center{
\includegraphics[scale=0.25]{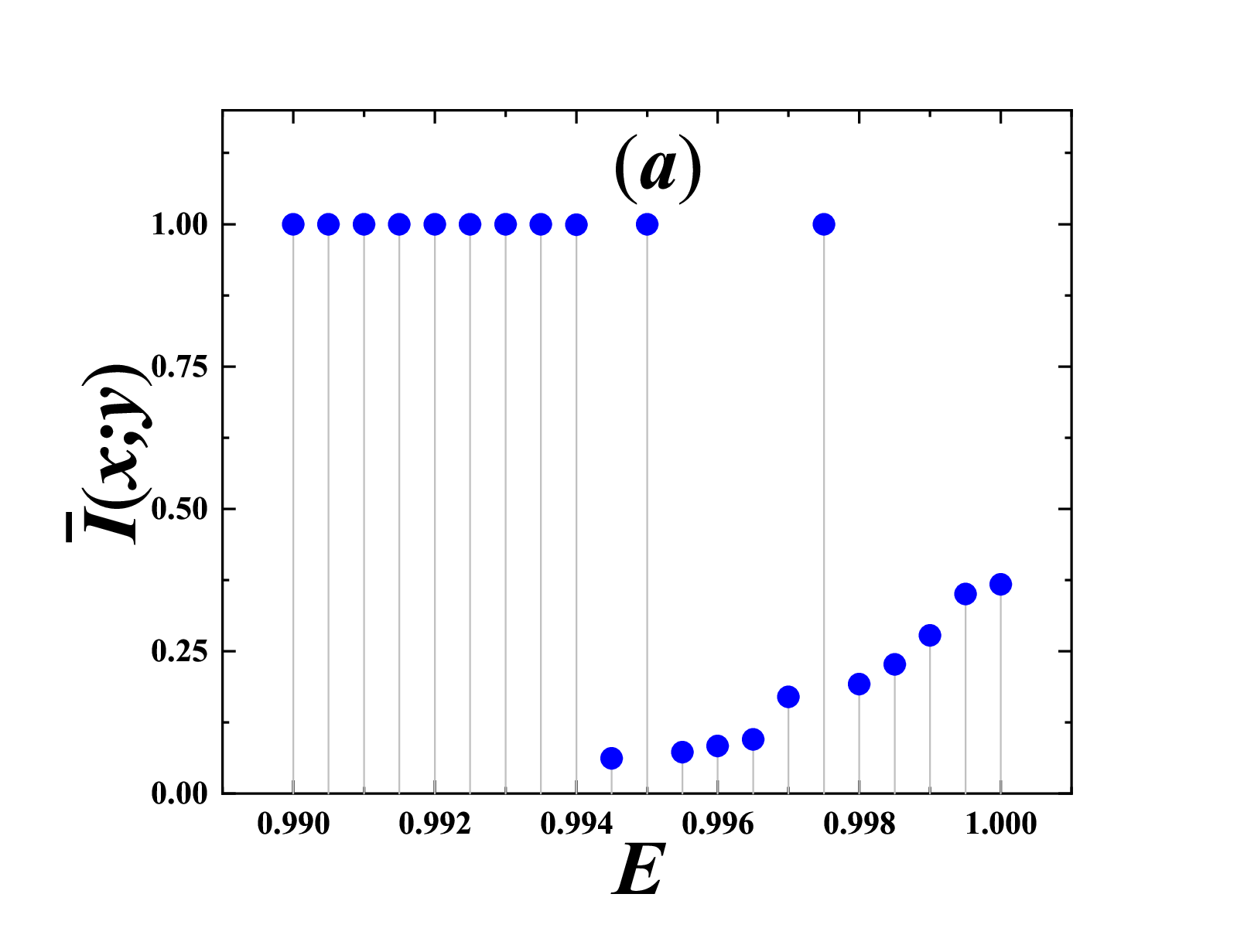}
\includegraphics[scale=0.25]{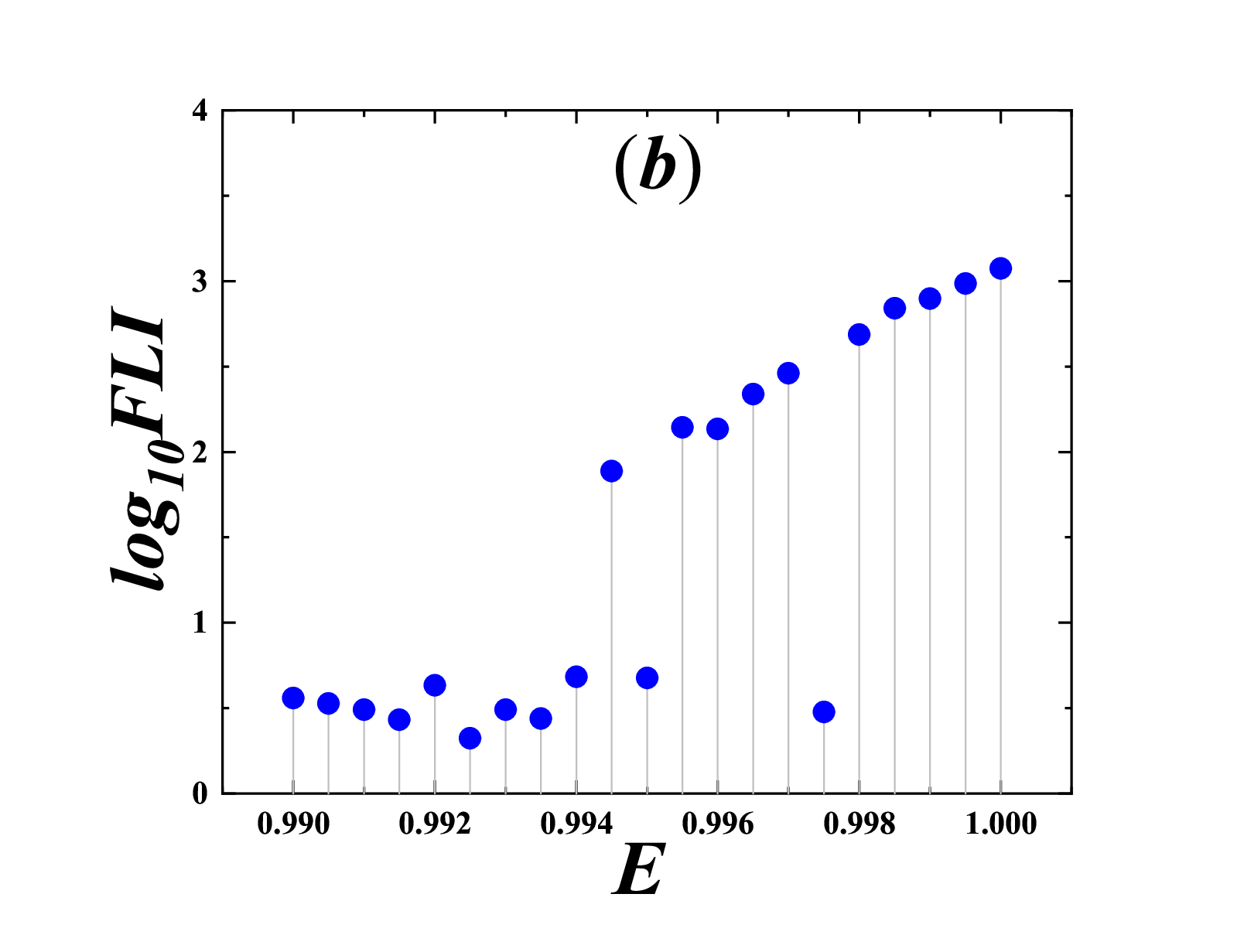}
\caption{MIPP(Cpu cost: 384s) and FLI(Cpu cost: 409s) scan plots for the particle energy $E$. The other orbital parameters are: $r=70$, $L=4.6$, $a=0.5$, $b=0.001$, and
$\theta=\frac{\pi}{2}$. A total of 21 orbits were studied.}
 \label{Fig5}}
\end{figure*}

\begin{figure*}[htbp]
\center{
\includegraphics[scale=0.25]{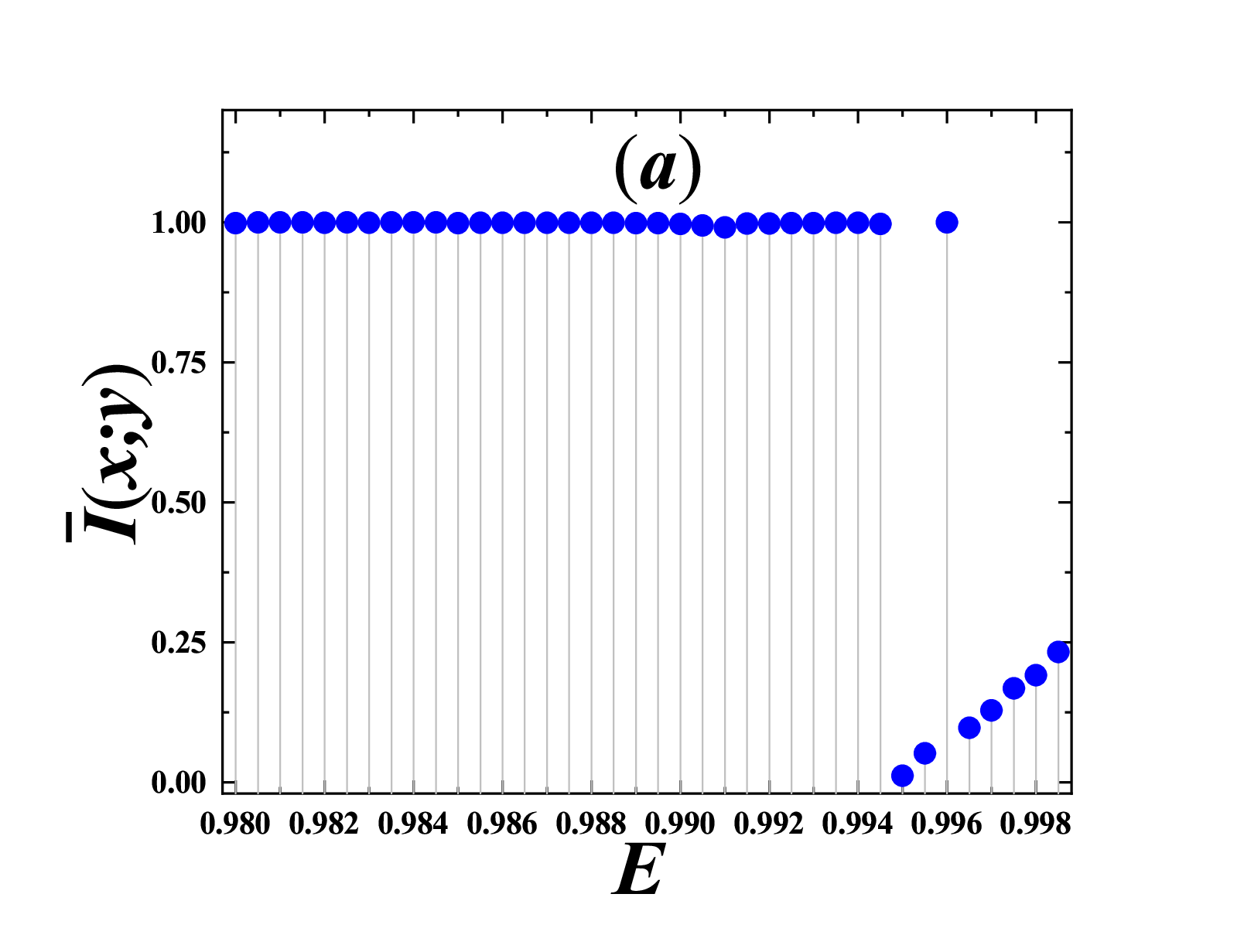}
\includegraphics[scale=0.25]{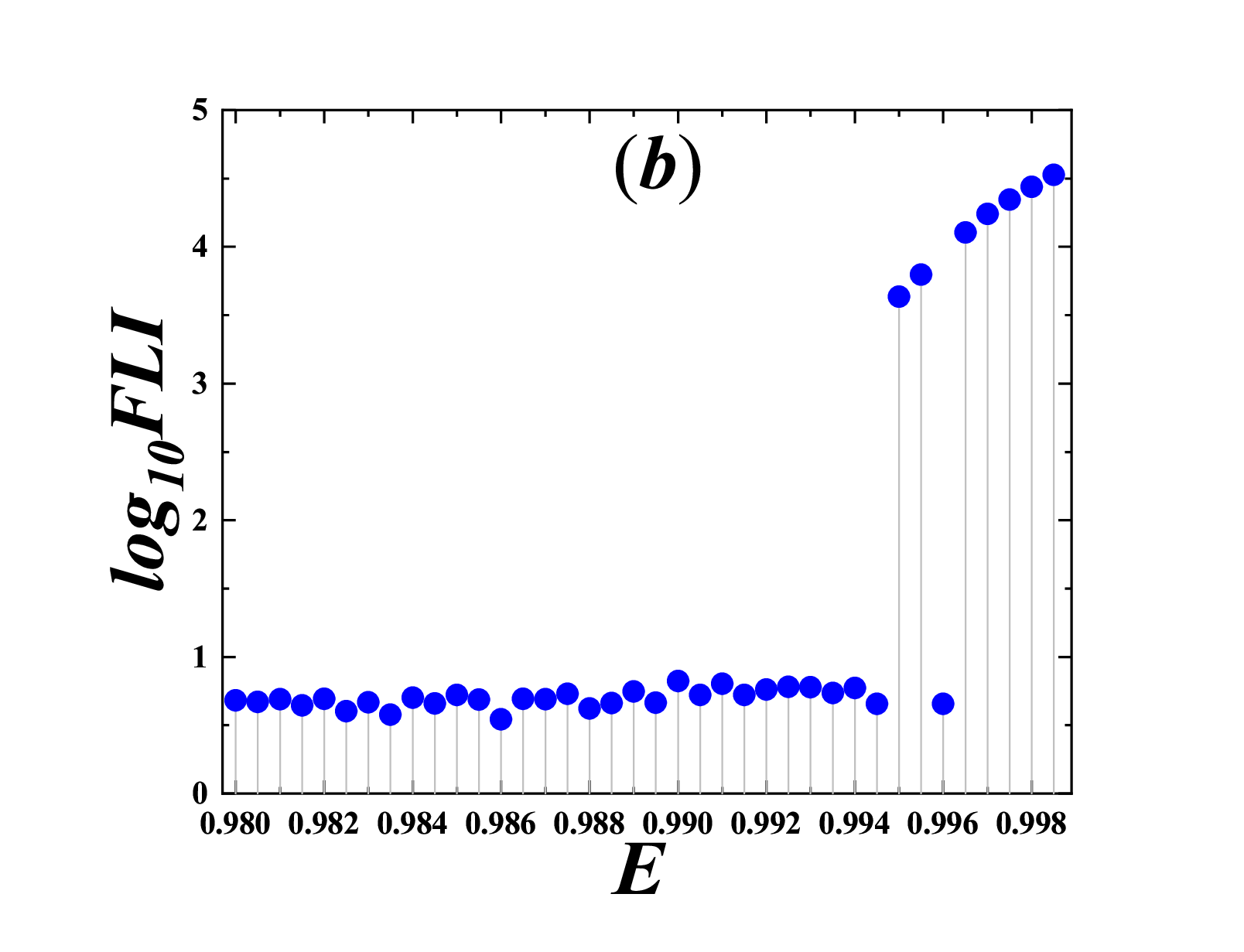}
\caption{MIPP(Cpu cost: 555s) and FLI(Cpu cost: 563s) scan plots for the energy E of the Schwarzschild black hole. The other orbital parameters are:  $L=4.6$, $r=11$,
$b=0.001$, and
$\theta=\frac{\pi}{2}$. A total of 38 orbits were studied.}
 \label{Fig6}}
\end{figure*}

\begin{figure*}[htbp]
\center{
\includegraphics[scale=0.25]{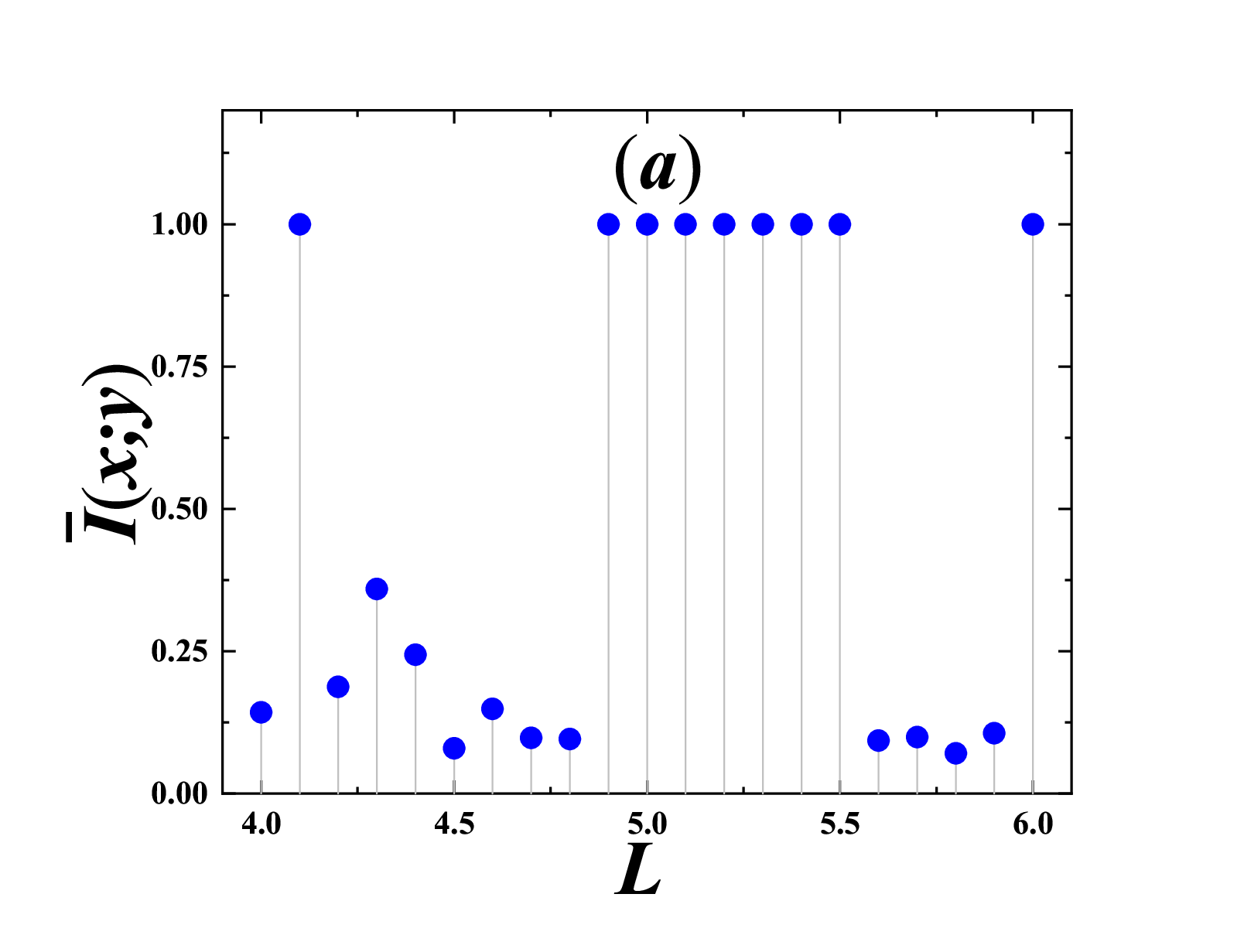}
\includegraphics[scale=0.25]{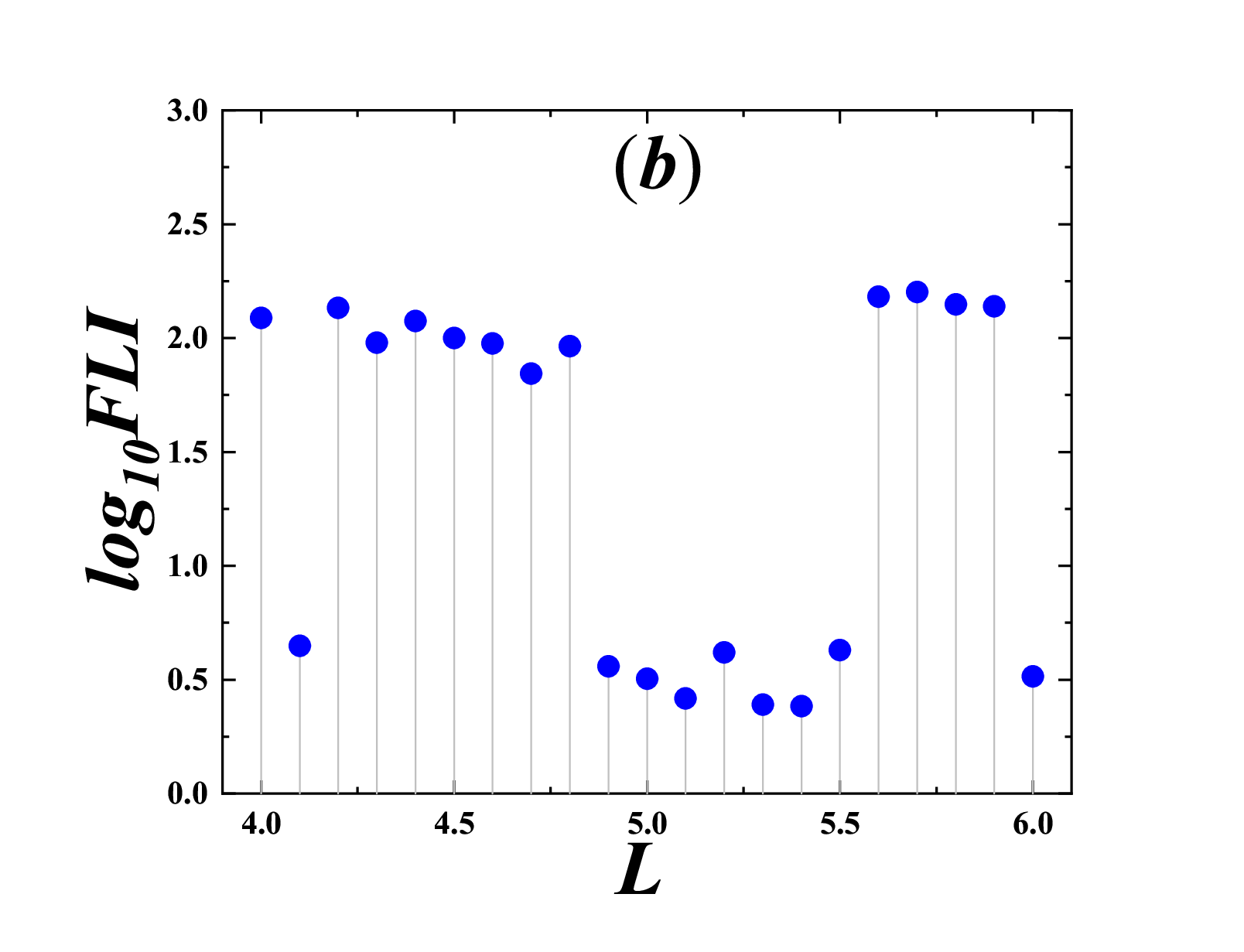}
\caption{MIPP(Cpu cost: 395s) and FLI(Cpu cost: 399s)  scan plots for the particle angular momentum $L$. The other orbital parameters are: $E=0.996$, $r=75$, $a=0.5$,
$b=0.001$, and $\theta=\frac{\pi}{2}$. A total of 21 orbits were studied.}
 \label{Fig7}}
\end{figure*}

\begin{figure*}[htbp]
\center{
\includegraphics[scale=0.25]{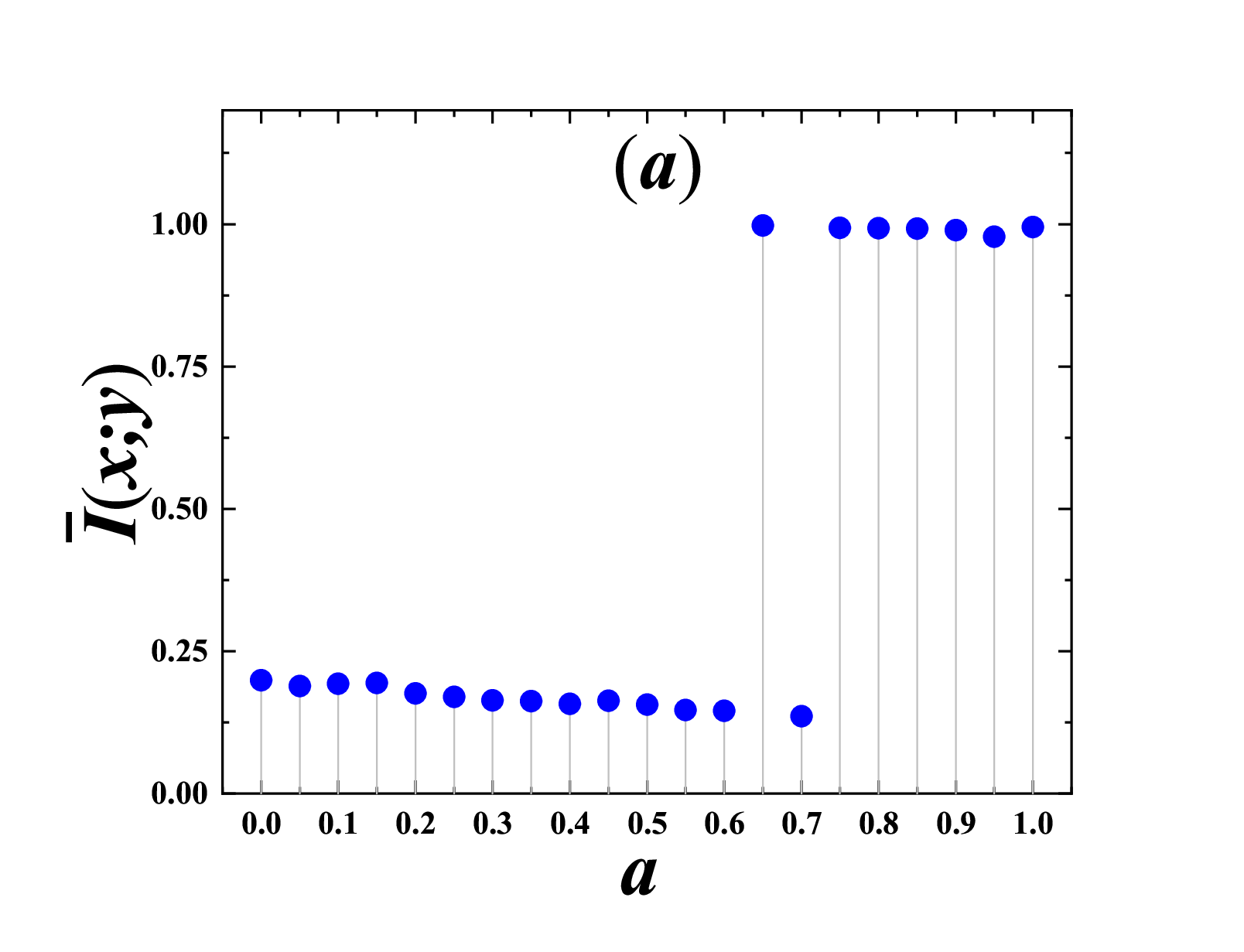}
\includegraphics[scale=0.25]{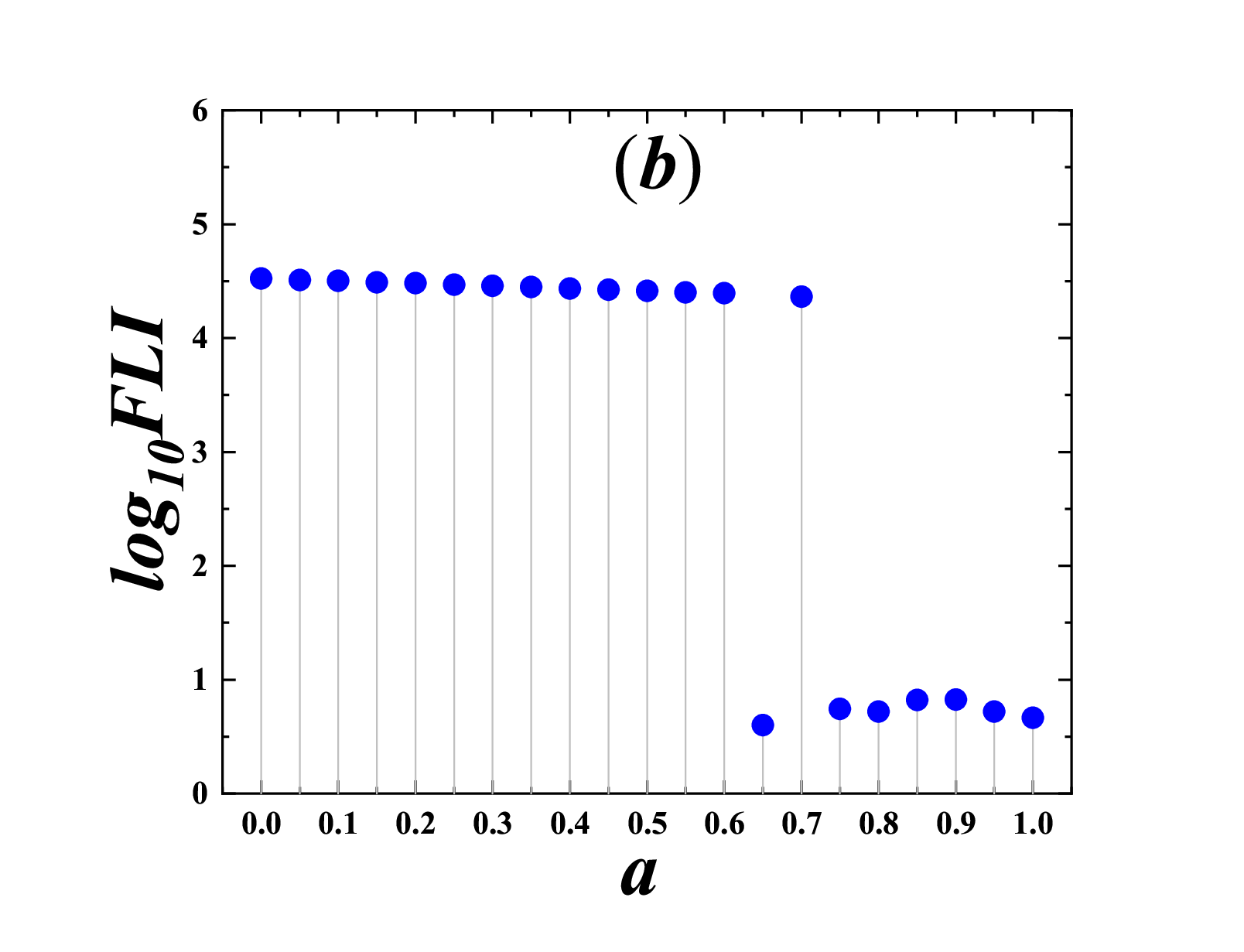}
\caption{MIPP(Cpu cost: 398s) and FLI(Cpu cost: 404s) scan plots for the black hole spin $a$. The other orbital parameters are: $E=0.998$, $L=4.6$, $r=10$, $b=0.001$, and
$\theta=\frac{\pi}{2}$. A total of 21 orbits were studied.}
 \label{Fig8}}
\end{figure*}


\begin{thebibliography}{99}


\bibitem{Devaney:1990}
Robert Devaney,
``An Introduction To Chaotic Dynamics Systems,''
The Mathematical Gazette \textbf{74} (1990).
doi:10.2307/3619398

\bibitem{Cao:2024bjk}
W.~Cao, Y.~Huang and H.~Zhang,
``Screen chaotic motion by Shannon entropy in curved spacetimes,''
[arXiv:2410.20870 [gr-qc]].


\bibitem{Lorenz:1963yb}
E.~N.~Lorenz,
``Deterministic nonperiodic flow,''
J. Atmos. Sci. \textbf{20}, 130-141 (1963)
doi:10.1175/1520-0469

\bibitem{Henon1964}
M.~Henon and C.~Heiles,
``The applicability of the third integral of motion: Some numerical experiments,''
Astronomical Journal. \textbf{69}, pp.~73, (1964).
doi: 10.1086/109234.

\bibitem{Gerald:1988}
Gerald Jay Sussman, Jack Wisdom,
``Numerical Evidence That the Motion of Pluto Is Chaotic,''
Science \textbf{241},433-437(1988).
DOI:10.1126/science.241.4864.433

\bibitem{Kopacek:2010yr}
O.~Kopacek, V.~Karas, J.~Kovar and Z.~Stuchlik,
``Transition from Regular to Chaotic Circulation in Magnetized Coronae near Compact Objects,''
Astrophys. J. \textbf{722}, 1240-1259 (2010)
doi:10.1088/0004-637X/722/2/1240
[arXiv:1008.4650 [astro-ph.HE]].

\bibitem{Chen:2016tmr}
S.~Chen, M.~Wang and J.~Jing,
``Chaotic motion of particles in the accelerating and rotating black holes spacetime,''
JHEP \textbf{09}, 082 (2016)
doi:10.1007/JHEP09(2016)082
[arXiv:1604.02785 [gr-qc]].

\bibitem{Li:2018wtz}
D.~Li and X.~Wu,
``Chaotic motion of neutral and charged particles in a magnetized Ernst-Schwarzschild spacetime,''
Eur. Phys. J. Plus \textbf{134}, no.3, 96 (2019)
doi:10.1140/epjp/i2019-12502-9
[arXiv:1803.02119 [gr-qc]].


\bibitem{Yi:2020shw}
M.~Yi and X.~Wu,
``Dynamics of charged particles around a magnetically deformed Schwarzschild black hole,''
Phys. Scripta \textbf{95}, no.8, 085008 (2020)
doi:10.1088/1402-4896/aba4c2

\bibitem{Hu:2021gwd}
A.~R.~Hu and G.~Q.~Huang,
``Dynamics of charged particles in the magnetized $\gamma $ spacetime,''
Eur. Phys. J. Plus \textbf{136}, no.12, 1210 (2021)
doi:10.1140/epjp/s13360-021-02194-1

\bibitem{Sun:2021oxg}
W.~Sun, Y.~Wang, F.~Liu and X.~Wu,
``Applying explicit symplectic integrator to study chaos of charged particles around magnetized Kerr black hole,''
Eur. Phys. J. C \textbf{81}, no.9, 785 (2021)
doi:10.1140/epjc/s10052-021-09579-7
[arXiv:2109.02295 [gr-qc]].

\bibitem{Sun:2021ndd}
X.~Sun, X.~Wu, Y.~Wang, C.~Deng, B.~Liu and E.~Liang,
``Dynamics of Charged Particles Moving around Kerr Black Hole with Inductive Charge and External Magnetic Field,''
Universe \textbf{7}, no.11, 410 (2021)
doi:10.3390/universe7110410
[arXiv:2111.04900 [gr-qc]].

\bibitem{Zhang:2023lrt}
L.~Zhang, S.~Chen, Q.~Pan and J.~Jing,
``Chaotic motion of scalar particle coupling to Chern\textendash{}Simons invariant in the stationary axisymmetric Einstein\textendash{}Maxwell dilaton black hole spacetime,''
Eur. Phys. J. C \textbf{83}, no.9, 828 (2023)
doi:10.1140/epjc/s10052-023-12008-6
[arXiv:2309.12604 [gr-qc]].

\bibitem{Cao:2024ihv}
W.~Cao, X.~Wu and J.~Lyu,
``Electromagnetic field and chaotic charged-particle motion around hairy black holes in Horndeski gravity,''
Eur. Phys. J. C \textbf{84}, no.4, 435 (2024)
doi:10.1140/epjc/s10052-024-12804-8
[arXiv:2404.19225 [gr-qc]].


\bibitem{Destounis:2021mqv}
K.~Destounis, A.~G.~Suvorov and K.~D.~Kokkotas,
``Gravitational-wave glitches in chaotic extreme-mass-ratio inspirals,''
Phys. Rev. Lett. \textbf{126}, no.14, 141102 (2021)
doi:10.1103/PhysRevLett.126.141102
[arXiv:2103.05643 [gr-qc]].


\bibitem{Tancredi:2001}
G.~Tancredi, A.~S\'{a}nchez, and F.~Roig,
``A Comparison Between Methods to Compute Lyapunov Exponents,''
Astron. J. \textbf{121}, no.2, 1171-1179 (2001)
doi:10.1086/318732.


\bibitem{Wu:2003pe}
X.~Wu and T.~y.~Huang,
``Computation of Lyapunov exponents in general relativity,''
Phys. Lett. A \textbf{313}, 77-81 (2003)
doi:10.1016/S0375-9601(03)00720-5
[arXiv:gr-qc/0302118 [gr-qc]].

\bibitem{Voglis1994}
N.~Voglis and G.~J.~Contopoulos,
``Invariant spectra of orbits in dynamical systems,''
Journal of Physics A: Mathematical and General, \textbf{27}, no.~14, pp.~4899, 1994.
DOI: 10.1088/0305-4470/27/14/017


\bibitem{Wu:2006rx}
X.~Wu, T.~Y.~Huang and H.~Zhang,
``Lyapunov indices with two nearby trajectories in a curved spacetime,''
Phys. Rev. D \textbf{74}, 083001 (2006)
doi:10.1103/PhysRevD.74.083001
[arXiv:1006.5251 [gr-qc]].

\bibitem{Binney1984}
J.~J. Binney and D. Spergel,
``Spectral stellar dynamics,''
Astrophys. J. \textbf{252}, 308-321 (1984),
doi:10.1086/159559.


\bibitem{Shannon:1948}
C. E. Shannon,
``A mathematical theory of communication,''
The Bell System Technical Journal \textbf{27}, no. 3, pp. 379-423(1948)
doi: 10.1002/j.1538-7305.1948.tb01338.x.


\bibitem{Eckmann:1987}
Eckmann, J. -P. \textit{et al.}
``Recurrence plots of dynamical systems,''
EPL (Europhysics Letters) \textbf{4},
pp. 973 (1987),
doi: 10.1209/0295-5075/4/9/004.

\bibitem{Gottwald:2009}
Gottwald, G. A. \textit{and} Melbourne, I.,
``On the Implementation of the 0--1 Test for Chaos,''
SIAM Journal on Applied Dynamical Systems \textbf{8}, no. 1,
pp. 129--145 (2009),
doi: 10.1137/080718851.

\bibitem{Parrondo:2015}
Parrondo, J., Horowitz, J., Sagawa, T.,
``Thermodynamics of information,''
Nature Phys \textbf{11}, 131-139 (2015)
https://doi.org/10.1038/nphys3230.

\bibitem{Cai:2017ihd}
R.~G.~Cai, X.~X.~Zeng and H.~Q.~Zhang,
``Influence of inhomogeneities on holographic mutual information and butterfly effect,''
JHEP \textbf{07}, 082 (2017)
doi:10.1007/JHEP07(2017)082
[arXiv:1704.03989 [hep-th]].

\bibitem{Carter:1968ks}
B.~Carter,
``Hamilton-Jacobi and Schrodinger separable solutions of Einstein's equations,''
Commun. Math. Phys. \textbf{10}, no.4, 280-310 (1968)
doi:10.1007/BF03399503


\bibitem{Cao:2022bvu}
W.~Cao, W.~Liu and X.~Wu,
``Integrability of Kerr-Newman spacetime with cloud strings, quintessence and electromagnetic field,''
Phys. Rev. D \textbf{105}, no.12, 124039 (2022)
doi:10.1103/PhysRevD.105.124039
[arXiv:2206.09518 [gr-qc]].


\bibitem{Wald:1974np}
R.~M.~Wald,
``Black hole in a uniform magnetic field,''
Phys. Rev. D \textbf{10}, 1680-1685 (1974)
doi:10.1103/PhysRevD.10.1680


\bibitem{Wang:2021gja}
Y.~Wang, W.~Sun, F.~Liu and X.~Wu,
``Construction of Explicit Symplectic Integrators in General Relativity. I. Schwarzschild Black Holes,''
Astrophys. J. \textbf{907}, no.2, 66 (2021)
doi:10.3847/1538-4357/abcb8d
[arXiv:2102.00373 [gr-qc]].

\bibitem{Wang:2021xww}
Y.~Wang, W.~Sun, F.~Liu and X.~Wu,
``Construction of Explicit Symplectic Integrators in General Relativity. II. Reissner\textendash{}Nordstr\"om Black Holes,''
Astrophys. J. \textbf{909}, no.1, 22 (2021)
doi:10.3847/1538-4357/abd701
[arXiv:2103.02864 [gr-qc]].


\bibitem{Wang:2021yqk}
Y.~Wang, W.~Sun, F.~Liu and X.~Wu,
``Construction of Explicit Symplectic Integrators in General Relativity. III. Reissner\textendash{}Nordstr\"om-(anti)-de Sitter Black Holes,''
Astrophys. J. Suppl. \textbf{254}, no.1, 8 (2021)
doi:10.3847/1538-4365/abf116
[arXiv:2103.12272 [gr-qc]].


\bibitem{Wu:2021rrd}
X.~Wu, Y.~Wang, W.~Sun and F.~Liu,
``Construction of Explicit Symplectic Integrators in General Relativity. IV. Kerr Black Holes,''
Astrophys. J. \textbf{914}, no.1, 63 (2021)
doi:10.3847/1538-4357/abfc45
[arXiv:2106.12356 [gr-qc]].

\bibitem{Wu:2022nye}
X.~Wu, Y.~Wang, W.~Sun, F.~Y.~Liu and W.~B.~Han,
``Explicit Symplectic Methods in Black Hole Spacetimes,''
Astrophys. J. \textbf{940}, no.2, 166 (2022)
doi:10.3847/1538-4357/ac9c5d
[arXiv:2210.13185 [gr-qc]].


\bibitem{Zhou:2022uht}
N.~Zhou, H.~Zhang, W.~Liu and X.~Wu,
``A Note on the Construction of Explicit Symplectic Integrators for Schwarzschild Spacetimes,''
Astrophys. J. \textbf{927}, no.2, 160 (2022)
[erratum: Astrophys. J. \textbf{947}, no.2, 94 (2023)]
doi:10.3847/1538-4357/ac497f
[arXiv:2201.02922 [gr-qc]].


\bibitem{Wu:2024ehd}
X.~Wu, Y.~Wang, W.~Sun, F.~Liu and D.~Ma,
``Explicit Symplectic Integrators with Adaptive Time Steps in Curved Spacetimes,''
Astrophys. J. Suppl. \textbf{275}, no.2, 31 (2024)
doi:10.3847/1538-4365/ad8351
[arXiv:2412.01045 [gr-qc]].

\bibitem{Liang:2023}
Canbin Liang, Bin Zhou,
``Differential Geometry and General Relativity. Volume 1,''
Translated by Weizhen Jia, Bin Zhou,
Graduate Texts in Physics,
Springer Singapore, 2023.
doi:10.1007/978-981-99-0022-0.



\end{thebibliography}
\end{document}